\documentclass[prd,twocolumn]{revtex4}

\pdfoutput=1

\usepackage{graphicx}
\usepackage{dcolumn}
\usepackage{bm}
\usepackage{amssymb,amsmath,bm}  
\usepackage{color}
\usepackage{hyperref}
\usepackage{multirow}
\usepackage[utf8]{inputenc}
\usepackage{balance}
\usepackage{enumitem}
\newcommand{\egr}{\epsilon_{\rm GR}}
\newcommand{\ewl}{\epsilon_{\rm WL}}
\newcommand{\fnl}{f_{\rm NL}}
\newcommand{\fevo}{f_{\rm evo}}
\newcommand{\uJy}{\mu{\rm Jy}}
\newcommand{\nv}{\hat{\bf n}}
\newcommand{\Tr}{{\rm Tr}}
\definecolor{calpha}{RGB}{255,51,51}
\definecolor{cbeta}{RGB}{51,51,255}
\newcommand{\calpha}{\textcolor{calpha}{1}}
\newcommand{\cbeta}{\textcolor{cbeta}{2}}

\begin{document}
\title{Constraining ultra large-scale cosmology with multiple tracers in optical and radio surveys}
\author{D. Alonso$^1$, P. G. Ferreira$^1$}
\affiliation{$^{1}$University of Oxford, Denys Wilkinson Building,
  Keble Road, Oxford, OX1 3RH,  UK}

\begin{abstract}
  Multiple tracers of the cosmic density field, with different bias, number and luminosity
  evolution, can be used to  measure the large-scale properties of the
  Universe. We show how an optimal combination of tracers can be used to detect
  general-relativistic effects in the observed density of sources. We forecast
  for the detectability of these effects, as well as measurements of primordial non-Gaussianity
  and large-scale lensing magnification with current and upcoming large-scale structure
  experiments. In particular we quantify the significance of these detections in the short
  term with experiments such as the Dark Energy Survey (DES), and in the long term with the
  Large Synoptic Survey Telescope (LSST) and the Square Kilometre Array (SKA). We review the
  main observational challenges that must be overcome to  carry out these measurements.
\end{abstract}

  \date{\today}
  \maketitle

\section{Introduction}\label{sec:intro}
  Over the next few decades we expect to map out the large scale structure of the observable
  Universe in its entirety. By this, we mean that we should  be able to completely
  characterize the spatial distribution of matter from sub-Megaparsec scales out to the
  cosmological horizon, from the present epoch out to the highest observable redshift, deep
  into the dark ages of structure formation. We will achieve this through a combination of
  redshift (both spectroscopic and photometric) and weak lensing surveys allied with intensity
  mapping and CMB experiments. From the spatial distribution of matter, we should be able to
  infer the structure of space-time within our cosmic horizon and in doing so dramatically
  improve our understanding of the cosmological model and its fundamental assumptions.
 
  Key to our understanding of the Universe are the very largest scales. There we expect to be
  truly sensitive to primordial and general-relativistic aspects of cosmology  and, one hopes,
  we should be able to constrain aspects of the Universe which are not accessible if we
  restrict ourselves to smaller, Newtonian, scales. The most notable aspects, which we will
  focus on in this paper and on which we will elaborate below, are general-relativistic
  effects on number counts and luminosities \cite{Bonvin:2005ps,2009PhRvD..80h3514Y,
  2010PhRvD..82h3508Y,2011PhRvD..84f3505B,2011PhRvD..84d3516C}, large-scale weak lensing and
  the scale dependence of bias due to primordial non-Gaussianity \cite{2008PhRvD..77l3514D,
  Matarrese:2008nc}.

  There have been  attempts at constraining the largest scales of the Universe with current
  surveys \cite{Giannantonio:2013uqa} and of quantifying how well we will be able to do so
  with future experiments \cite{2011PhRvD..83l3514N, 2012MNRAS.422.2854G, 2013PhRvL.111q1302C, 
  2014MNRAS.442.2511F, 2015MNRAS.448.1035C, 2015JCAP...01..042R}.  Most of these attempts have
  focused on autocorrelations of single tracers, where it has become clear that cosmic (or
  sample) variance is sufficiently large that the resulting constraints are not (or will not be)
  particularly stringent. In other words, the fundamentally stochastic nature of the seeds for
  structure combined with the finite volume of any survey severely restricts the number of
  independent modes we can sample and introduces a large and inescapable floor on the
  uncertainty of any of the fundamental quantities we wish to constrain. In
  \cite{2015arXiv150507596A} we undertook a detailed, comprehensive analysis of the four
  different types of surveys of large scale structure which are currently being planned, finding
  indeed that there were severe limitations to how well we could constrain large scale effects.

  An ingenious and alternative approach can be taken if one has multiple tracers (such as
  galaxies or line-emitting species) with different bias and luminosity functions. As was first
  shown in \cite{2009PhRvL.102b1302S} and then further developed in \cite{Abramo:2013awa}, with
  multiple tracers it is possible to effectively "divide out" the stochastic dependence of the
  cosmological density field allowing us to beat cosmic variance for a few, specific,
  observables such as the growth rate and clustering bias. While this approach
  was originally advocated to target the scale dependent bias of primordial non-Gaussianity
  \cite{2009PhRvL.102b1302S}, we expect that such an effect naturally extends to the
  general-relativistic corrections in large-scale structure surveys.

  In this paper, we further flesh out the idea of using multi-tracer methods to constrain what
  we dub {\it ultra large-scale} effects. We focus on surveys that have two key properties.
  First, they must be able to cover large volumes of space with an appreciable number density of
  sources. Second, we need to be able to pick out two or more different tracers with a
  sufficiently high bias but also with sufficiently different evolution properties that the
  multi-tracer technique can be effective. In subsection \ref{ssec:multi_why} we derive a
  simple expression which can be used as a guide for what the best combinations of tracers
  (and their properties) are to efficiently constrain general relativistic effects. Both of these
  requirements lead us to choose photometric redshift surveys (focusing on DES and LSST),
  radio-continuum surveys and intensity mapping surveys (primarily from the SKA). Indeed we
  find that a combination of these surveys will give us remarkable constraints on ultra
  large-scale effects.

  We structure our paper as follows. In Section \ref{sec:th_uls} we remind ourselves of the
  various effects that come into play on large scales, explaining their origin and how they
  affect the observables we will consider. In Section \ref{sec:th_fisher} we describe the
  Fisher forecasting formalism we will be using and employ it to derive a simple expression which
  shows how the multi-tracer method works to our advantage, for isolating the general-relativistic
  corrections to large-scale surveys. This expression gives us a few pointers to
  what the optimal surveys will be for us to achieve our goals. In Section \ref{sec:photo} we
  focus on optical and near infra-red photometric redshift surveys, specifically on LSST, in
  Section \ref{sec:radio} we look at cosmological radio surveys, specifically continuum and
  intensity mapping as will be carried out by the SKA, and in Section \ref{sec:im_vs_photo} we
  will see how cross-correlating these different surveys will contribute to our overall goals.
  Our results are truly promising and in Section  \ref{sec:discussion} their limitations and future
  prospects are presented. In particular we speculate on how precise measurements on ultra-large   
  scales will contribute to our understanding of the fundamental cosmological model as well as
  how they might help us further characterize the large scale properties of the Universe.

\section{Cosmological observables on ultra-large scales}\label{sec:th_uls}
  The aim of this section is to introduce the two main observables on ultra-large
  scales that we will study here, as well as to establish the notation that
  will be used in the rest of the paper, which follows the one used in
  \cite{2015arXiv150507596A}. More thorough discussions of these topics can be
  found in, e.g. \cite{2011PhRvD..84d3516C} for relativistic effects and
  \cite{Matarrese:2008nc} for primordial non-Gaussianity.
  
  \subsection{Relativistic effects in large-scale structure}\label{ssec:th_rel}
    The aim of large-scale structure experiments is to map out the three-dimensional
    distribution of the matter density field in the Universe. This is most commonly
    done using the number counts of astrophysical sources in the sky, using their
    angular positions and redshifts as proxies for their three-dimensional coordinates.
    This means that we use the redshift and direction of the photons emitted by
    these sources in order to infer the comoving volume of the patch over which they
    are distributed; any perturbation in the trajectory or the frequency
    of these photons along the lightcone, will affect our volume estimates, and
    consequently the measured three-dimensional density map.
    
    Two perfect examples of
    such effects are redshift-space distortions (RSDs), by which the emitted photons
    are further redshifted by the peculiar velocities of the sources, and
    gravitational lensing magnification, the perturbation in the trajectory of the
    photons caused by the gravitational field of the intervening matter. RSDs displace
    the sources in redshift-space coherently along the line of sight, while lensing
    magnification modifies the solid angle subtended by a cluster of sources.
    While these two effects correspond to the two largest contributions to the total
    fluctuation in the observed number counts, after the intrinsic perturbation in
    the source number density, there are other non-zero contributions at linear order
    that potentially contain information about the nature of the gravitational
    interactions perturbing the photon trajectories. The complete set of terms
    at linear order was presented by \cite{2011PhRvD..84d3516C} and
    \cite{2011PhRvD..84f3505B}, and can be classified into four main contributions:
    \begin{equation}\label{eq:pert1}
      \Delta_N(z,\nv)=\delta_{\mathcal{N}}+
      \frac{\partial\ln\bar{\mathcal{N}}}{\partial\eta}\delta\eta+
      \delta_\parallel+2\delta_\perp\,
      \left[1-\frac{\partial\ln\bar{\mathcal{N}}}{\partial \ln L_*}\right],
    \end{equation}
    where $\bar{\mathcal{N}}(\eta,\ln L_*)$ is the physical number density of sources
    at conformal time $\eta$ with luminosities above $L_*$ (i.e. the cumulative
    luminosity function), related to the survey's flux limit $F_{\rm cut}$ via
    \begin{equation}
      L_*(z)=4\pi(1+z)^2\chi^2(z)\,F_{\rm cut}.
    \end{equation}
    The four terms on the right hand side of Eq. \ref{eq:pert1} correspond to the
    intrinsic perturbation in the number density of sources ($\delta_\mathcal{N}$),
    the difference between the source's conformal time $\eta$ and the conformal
    time inferred from the redshift $\eta_*$ ($\delta\eta$), and the corresponding
    differences in the radial distance ($\delta_\parallel$) and angular diameter
    distance ($\delta_\perp$) of the sources. As discussed above, RSDs and lensing
    magnification contribute to $\delta_\parallel$ and $\delta_\perp$ respectively.
    In a given gauge, we can expand each of these in terms of the metric, density
    and velocity perturbations. This has been done explicitly in the literature,
    and we present  expressions for the complete set of terms in Appendix
    \ref{app:full_expressions}.
    
    Note that the amplitude of the observed perturbation in the number counts as
    a function of redshift depends on three tracer-specific quantities:
    \begin{itemize}
      \item The \emph{clustering bias} $b$, relating the intrinsic perturbation in
            the number density with the fluctuations in the comoving-gauge matter
            perturbation, defined in Fourier space as \cite{2011PhRvD..84d3516C,
            Bruni:2011ta,2011JCAP...10..031B}:
            \begin{equation}
              \delta_\mathcal{N}(k)=b\,\delta_M^{\rm synch}+
              \frac{\partial\ln\bar{\mathcal{N}}}{\partial\eta}\frac{v}{k},
            \end{equation}
            where $\delta_M^{\rm synch}$ is the matter perturbation in the synchronous
            comoving gauge and $v$ is the peculiar velocity in the Newtonian gauge.
      \item The slope of the background number density of sources as a function of
            time, which is now customary to encode in the so-called \emph{evolution
            bias}:
            \begin{equation}
              \fevo\equiv\frac{\partial\ln(a^3\,\bar{\mathcal{N}})}{\partial\ln(a)}.
            \end{equation}
      \item The slope of the number density of sources as a function of intrinsic
            luminosity, usually written in term of the \emph{magnification bias}:
            \begin{equation}
              s\equiv\frac{5}{2}\frac{\partial\ln(\bar{\mathcal{N}})}{\partial\ln L_*}.
            \end{equation}
    \end{itemize}

    In general, the astrophysical observable containing the most cosmological
    information is the perturbation in the number counts in a set of redshift
    bins
    \begin{equation}
     \Delta_i(\nv)\equiv\int dz\,W_i(z)\Delta(z,\nv),
    \end{equation}
    where $W_i(z)$ is the the window function defining the $i-$th redshift bin
    (normalized to 1 when integrated over the whole redshift range). To make
    matters simpler, we will present our forecasts in terms of the harmonic
    coefficients of $\Delta_i(\nv)$, defined as
    \begin{equation}
      a^i_{\ell m}\equiv\int d\nv\, \Delta_i(\nv)\,Y_{\ell m}(\nv).
    \end{equation}
    The main advantage of using the $a_{\ell m}$s is that statistical isotropy
    makes them uncorrelated, so that:
    \begin{equation}
      \langle a^i_{\ell m}a^j_{\ell' m'}\rangle\equiv C^{ij}_\ell\,
      \delta_{\ell\ell'}\delta_{mm'}.
    \end{equation}
    This equation also defines the angular power spectrum $C^{ij}_\ell$,
    containing all of the possible cross-correlations between two redshift
    bins $i$ and $j$.
    
    As shown in \cite{2013JCAP...11..044D}, the angular power spectrum can
    be computed as
    \begin{equation}\label{eq:th_cl}
      C^{ij}_\ell=4\pi\int_0^\infty\frac{dk}{k}\mathcal{P}(k)
      \,\Delta_\ell^i(k)\Delta_\ell^i(k),
    \end{equation}
    where $\mathcal{P}(k)$ is the dimensionless primordial power spectrum, which
    we assume to be close to scale invariant: $\mathcal{P}(k)=
    A_s(k/k_0)^{n_s-1}$, and $\Delta_\ell^i$ is a transfer function containing
    the terms in Eq. \ref{eq:pert1} projected along the radial direction in
    each redshift bin. In the Newtonian gauge $\Delta^i_\ell$ can be expanded
    as a sum of 10 individual terms,
    \begin{eqnarray}
      \Delta\equiv& \Delta^{\rm D}+\Delta^{\rm RSD}+\Delta^{\rm L}+
      \Delta^{\rm V1} +\Delta^{\rm V2}\nonumber \\&+
      \Delta^{\rm P1}+\Delta^{\rm P2}+\Delta^{\rm P3}+\Delta^{\rm P4}+\Delta^{\rm ISW},
    \end{eqnarray}    
    where the complete expressions for each term are presented in
    Appendix \ref{app:full_expressions}.

    The first three terms, $\Delta^{\rm D}$, $\Delta^{\rm RSD}$ and $\Delta^{\rm L}$
    are the dominant contributions to the total fluctuation, and correspond to the
    intrinsic perturbation in the source number density, the Kaiser-term
    \cite{1987MNRAS.227....1K} for redshift-space distortions and the effect of lensing
    magnification respectively. The remaining 7 terms are subdominant and have never been
    accounted for, nor detected in the analysis of any dataset. Of these, the terms
    $\Delta^{\rm V1}$ and $\Delta^{\rm V2}$ are extra RSD effects due to evaluating
    the background terms at the redshift perturbed by the Doppler effect,
    and the remaining terms correspond to similar effects caused by gravitational
    redshifting and lensing (e.g. the Shapiro time delay $\Delta^{\rm P4}$ or the
    integrated Sachs-Wolfe effect $\Delta^{\rm ISW}$ \cite{1964PhRvL..13..789S,
    1967ApJ...147...73S}). A great deal of the work presented in this paper is devoted to
    studying the detectability of these last 7 terms, and the potential benefits that such
    a detection could entail.

    The contribution of these terms to the observed fluctuations is most relevant
    on the scale of the cosmic horizon, and even there they are subdominant with respect to
    the first three. Moreover, since accurate clustering measurements on these 
    scales can be severely hampered by observational systematics, as we will review
    later, detecting the effect of these terms is very challenging. In particular,
    as was shown in \cite{2015arXiv150507596A}, the amplitude of these terms is too
    small to be observable with any significance using any of the main cosmological
    probes targeted by next-generation large-scale structure experiments individually.
    The main reason for this is the large cosmic variance on these scales, both
    due to the small number of super-horizon modes available at redshifts
    $z\lesssim3$ and to the larger amplitude of matter fluctuations at late
    times. In order to detect these relativistic corrections, it is therefore
    necessary to reduce the statistical noise below the cosmic variance limit,
    a feat that might be achievable using the multi-tracer techniques that we
    review in section \ref{ssec:fisher_multi}. We will quantify the detectability
    of these GR corrections as was done in \cite{2015arXiv150507596A}, by defining an
    effective parameter $\egr$ multiplying the terms $\Delta^{\rm V1,2}_\ell$,
    $\Delta^{\rm P1-4}_\ell$ and $\Delta^{\rm ISW}_\ell$. We do not include
    the lensing magnification term $\Delta^{\rm L}_\ell$ under the umbrella
    of $\egr$, since this effect has already been detected observationally
    \cite{2005ApJ...633..589S, 2009A&A...507..683H}, and its contribution is also
    non-negligible on small angular scales. However, for completenes, and given that
    gravitational lensing is clearly a General-Relativistic effect, we will also
    forecast for its detectability by defining a separate amplitude parameter for
    it, $\ewl$.

    To end this section, it is worth noting that, even though the form of the
    fluctuation $\Delta_i(\nv)$ has been motivated for experiments targeting a
    flux-limited population of discrete sources, it is straightforward to extend
    it to the case of line-emission tomography \cite{2013PhRvD..87f4026H} (most
    commonly called \emph{intensity mapping} ). The main difference between these two
    cases is that, while in a survey of discrete sources the proxy for the density
    perturbation is the fluctuation in the number of objects detected
    above a given flux in different patches of equal size, the observable in 
    intensity mapping is the combined emission from all of the sources in these
    patches. Since lensing effects perturb angular separations in the same way as
    source luminosities, the net effect is an exact cancellation of the linear-order
    perturbation in the angular diameter distance for intensity mapping (i.e. the
    $\delta_\perp$ term in Equation \ref{eq:pert1}). Thus, for the purposes of
    predicting the cosmological signal, intensity mapping experiments can be treated as
    a survey of discrete sources with a magnification bias given by the critical
    value $s_{\rm IM}=2/5$, and with an evolution bias given by:
    \begin{equation}\label{eq:fevo_im}
     f_{\rm evo}^{\rm IM}\equiv\frac{\partial\ln\rho_a}{\partial\ln a},
    \end{equation}
    where $\rho_a$ is the background comoving density of the line-emitting species
    under study.
  
  \subsection{Primordial non-Gaussianity}\label{ssec:th_png}
    One of the fundamental assumptions of the simplest versions of the standard
    cosmological model is an almost-negligible amount of non-Gaussianity in
    the statistics of the primordial density field \cite{Baumann:2009ds}. A simple (although not
    universal) formalism \cite{Komatsu:2001rj} to quantify deviations with respect to this premise
    is to decompose the primordial gravitational potential $\Phi$ into 
    a linear and a quadratic function of a Gaussian random field $\phi_G$:
    \begin{equation}
      \Phi=\phi_G+f_{\rm NL}\,(\phi_G^2-\langle\phi_G^2\rangle).
    \end{equation}
    
    Since the matter density field gradually becomes non-Gaussian due to the
    non-linear nature of gravitational collapse, the level of primordial
    non-Gaussianity is easier to determine using early-Universe probes, such
    as the CMB, which are well inside the linear regime. Current constraints
    based on higher-order CMB statistics favour a rather small level of
    non-Gaussianity, with $|\fnl|\lesssim7$ \cite{2015arXiv150201592P}. However, other
    statistical techniques have been developed with the aim of improving these
    constraints using low-redshift three-dimensional datasets. In particular
    it has been noted that a non-zero primordial non-Gaussianity would affect
    the formation and therefore the clustering statistics of biased tracers
    of the matter distribution \cite{2008PhRvD..77l3514D,Matarrese:2008nc}. The effect 
    can be encoded as a modification of the Gaussian clustering bias
    $b_G(z)\rightarrow b_G(z)+\Delta b(z,k)$, with a scale dependence given by
    \begin{equation}
      \Delta b(z,k)=3\fnl\frac{[b_G(z)-1]\Omega_M\,H_0^2\delta_c}{k^2T(k)D(z)},
    \end{equation}
    where $\delta_c\simeq1.686$ is the linearized collapse threshold, $T(k)$ is
    the matter transfer function and $D(z)$ is the linear growth factor.

    This scale-dependent bias induces extra power on large scales with
    a $\sim k^{-2}$ scale dependence, and it is easy to show that this
    extra power should be more evident on scales or the order of the horizon
    $k_{\rm NG}\sim\fnl\,H_0$. Thus this signature of primordial non-Gaussianity
    presents characteristics that are very similar to the signal of the GR
    corrections presented in the previous sections (in fact it has been
    suggested that they could potentially be degenerate \cite{Bruni:2011ta,
    Jeong:2011as,Bertacca:2012tp}), and therefore it makes sense to study both
    effects simultaneously.

    Including our three main observables ($\fnl$, $\egr$ and $\ewl$), the
    total number count fluctuation is:
    \begin{equation}
      \Delta_\ell=\Delta^{\rm D}_\ell(\fnl)+\Delta^{\rm RSD}_\ell+
      \ewl\,\Delta^{\rm L}_\ell+\egr\,\Delta^{\rm GR}_\ell,
    \end{equation}
    where every term is implicitly scale and time-dependent, and
    $\Delta^{\rm GR}\equiv\Delta^{\rm V1}+\Delta^{\rm V2}+
    \Delta^{\rm P1}+\Delta^{\rm P2}+\Delta^{\rm P3}+\Delta^{\rm P4}+
    \Delta^{\rm ISW}$.

\section{Forecasting formalism}\label{sec:th_fisher}
  The aim of this paper is to produce forecasts simultaneously for the detectability
  of the general relativistic corrections in terms of $\egr$ and for the level of
  primordial non-Gaussianity parametrized by $\fnl$. We do so using the Fisher matrix
  approach, extending the formalism that was used in \cite{2015arXiv150507596A} to
  include different tracers.

  \subsection{Fisher matrix with multiple tracers}\label{ssec:fisher_multi}
    Let us consider an experiment targeting a number of  tracers, each
    labelled by a Greek index $\alpha$; each tracer results in  a number of maps of the
    sky, labelled by a Roman index $i$. For the purposes of this paper, these maps
    correspond to different redshift bins or frequency bands. All the cosmological
    information is contained in the harmonic coefficients for each tracer and map, which
    we label by a combined index $(\alpha,i)$ (i.e. $a^{(\alpha,i)}_{\ell m}$ are the
    harmonic coefficients for the $\alpha$-th tracer in its $i$-th redshift bin). For
    Gaussian random fields, the  most important observable is the power spectrum,
    defined as the covariance of the $a^{(\alpha,i)}_{\ell m}$s:
    \begin{equation}
      \langle a^{(\alpha,i)}_{\ell m} (a^{(\beta,j)}_{\ell' m'})^*\rangle\equiv
      \delta_{\ell\ell'}\delta_{m m'}\,C^{(\alpha,i)(\beta,j)}_{\ell m}.
    \end{equation}
    
    We will assume that all tracers can be cleanly separated into signal and noise
    contributions $a^{(\alpha,i)}_{\ell m}=s^{(\alpha,i)}_{\ell m}+
    n^{(\alpha,i)}_{\ell m}$, and that both components are statistically
    independent. Furthermore, we will assume that the noise is uncorrelated between
    different tracers:
    \begin{equation}
      \langle n^{(\alpha,i)}_{\ell m} (n^{(\beta,j)}_{\ell' m'})^*\rangle\equiv
      N^{(\alpha),ij}_{\ell m}\delta_{\alpha\beta}\delta_{\ell\ell'}\delta_{mm'}.
    \end{equation}
    This assumption would not be correct, for instance, if the two tracers were two
    different but not disjoint galaxy populations. For most applications it is also
    safe to assume that the noise power spectrum is uncorrelated between different
    redshift bins.

    Arranging the combined indices $(\alpha,i)$ as a single one-dimensional index,
    we can write the $a^{(\alpha,i)}_{\ell m}$s and the power spectra as a vector
    ${\bf a}_{\ell m}$ and a matrix $\mathsf{C}_\ell$ respectively. For example, for
    two tracers $\alpha=\{1,2\}$ distributed over 2 and 1 redshift bins respectively,
    we would have ${\bf a}_{\ell m}=(a^{(\calpha,1)}_{\ell m},a^{(\calpha,2)}_{\ell m},
    a^{(\cbeta,1)}_{\ell m})$, and a power spectrum
    \begin{align}
      &\mathsf{C}_\ell=\left(\begin{array}{cc|c}
        C^{(\calpha,1)(\calpha,1)}_\ell &
        C^{(\calpha,1)(\calpha,2)}_\ell &
        C^{(\calpha,1)(\cbeta,1)}_\ell \\
        &
        C^{(\calpha,2)(\calpha,2)}_\ell &
        C^{(\calpha,2)(\cbeta,1)}_\ell \\
        &&\\\hline &&\\
        &
        &
        C^{(\cbeta,1)(\cbeta,1)}_\ell,
      \end{array}\right),
    \end{align}
    where we have omitted the redundant components strictly below the diagonal.
    It is then easy to show that the Fisher matrix for the observable ${\bf a}_{\ell m}$
    is the same as for the single tracer case, where the power spectrum matrix now contains
    also all possible cross-correlations between different tracers. Thus, for a set of
    parameters $\{\theta_\mu\}$, the covariance matrix $C_{\mu\nu}\equiv\langle
    (\theta_\mu-\bar{\theta}_\mu)(\theta_\nu-\bar{\theta}_\nu)\rangle$ can be
    approximated by the inverse of $F_{\mu\nu}$, where
    \begin{equation}\label{eq:fisher}
      F_{\mu\nu}=f_{\rm sky}\sum_{\ell=2}^{\ell_{\rm max}}\frac{2\ell+1}{2}
      \Tr\left[\partial_\mu\mathsf{C}_\ell\,\mathsf{C}_\ell^{-1}
        \partial_\nu\mathsf{C}_\ell\,\mathsf{C}_\ell^{-1}\right]
    \end{equation}
    and $\partial_\mu$ is the partial derivative with respect to the $\mu$-th
    parameter. In our forecasts the derivatives were computed using central finite differences,
    \begin{equation}
      \partial_\mu f=\frac{f(\theta_\mu+\delta\theta_\mu)-
        f(\theta_\mu-\delta\theta_\mu)}{2\delta\theta_\mu}+
      \mathcal{O}(\delta\theta^3),
    \end{equation}
    where the intervals $\delta\theta_\mu$ are such that the
    derivatives converge to the required numerical accuracy.

    As we pointed out above, we can divide the power spectrum up between two
    uncorrelated signal and noise contributions, $\mathsf{C}_\ell=
    \mathsf{C}^S_\ell+\mathsf{N}_\ell$. The cosmological power spectra
    $\mathsf{C}^S_\ell$ are given by Equation \ref{eq:th_cl}, and were calculated
    using a modified version of the public {\tt CLASS} code \cite{2011arXiv1104.2932L,
    2013JCAP...11..044D}. This version is based on the code used in
    \cite{2015arXiv150507596A}, with added capability to handle an arbitrary number
    of tracers \footnote{This code is made publicly available at
    \url{http://intensitymapping.physics.ox.ac.uk/codes.html}.}. For our
    fiducial cosmology we adopted a model consistent with the best-fit flat
    $\Lambda$CDM parameters from Planck \cite{2014A&A...571A..16P}, given
    by $(\Omega_M,f_b,h,w,A_s,n_s)=(0.315,0.156,0.67,-1,2.46\times10^{-9},0.96)$.
    We further set $\fnl=0$, the value for Gaussian initial conditions, and
    $\egr=\ewl=1$. In our forecasts we use these 9 free parameters, as well as
    a number of nuisance parameters associated with the bias functions of the
    tracers under study, which we describe below. Any priors on these parameters,
    correlated or not, were added as Gaussian priors to the bare Fisher matrix
    ($F_{\mu\nu}\rightarrow F_{\mu\nu}+(C^{\rm prior})^{-1}_{\mu\nu}$). In
    particular for the cosmological parameters listed above (excluding
    $\fnl,\,\egr$ and $\ewl$) we used the covariance matrix estimated from the
    corresponding Planck MCMC chains. We must also note that, since $\egr$ and
    $\ewl$ are only effective parameters, used to characterize the level of
    detection of the GR and lensing effects, when forecasting for $\fnl$ we
    fix both $\egr$ and $\ewl$ to their actual value of 1 instead of marginalizing
    over them.
    
    \begin{figure}
      \begin{center}
        \includegraphics[width=0.49\textwidth]{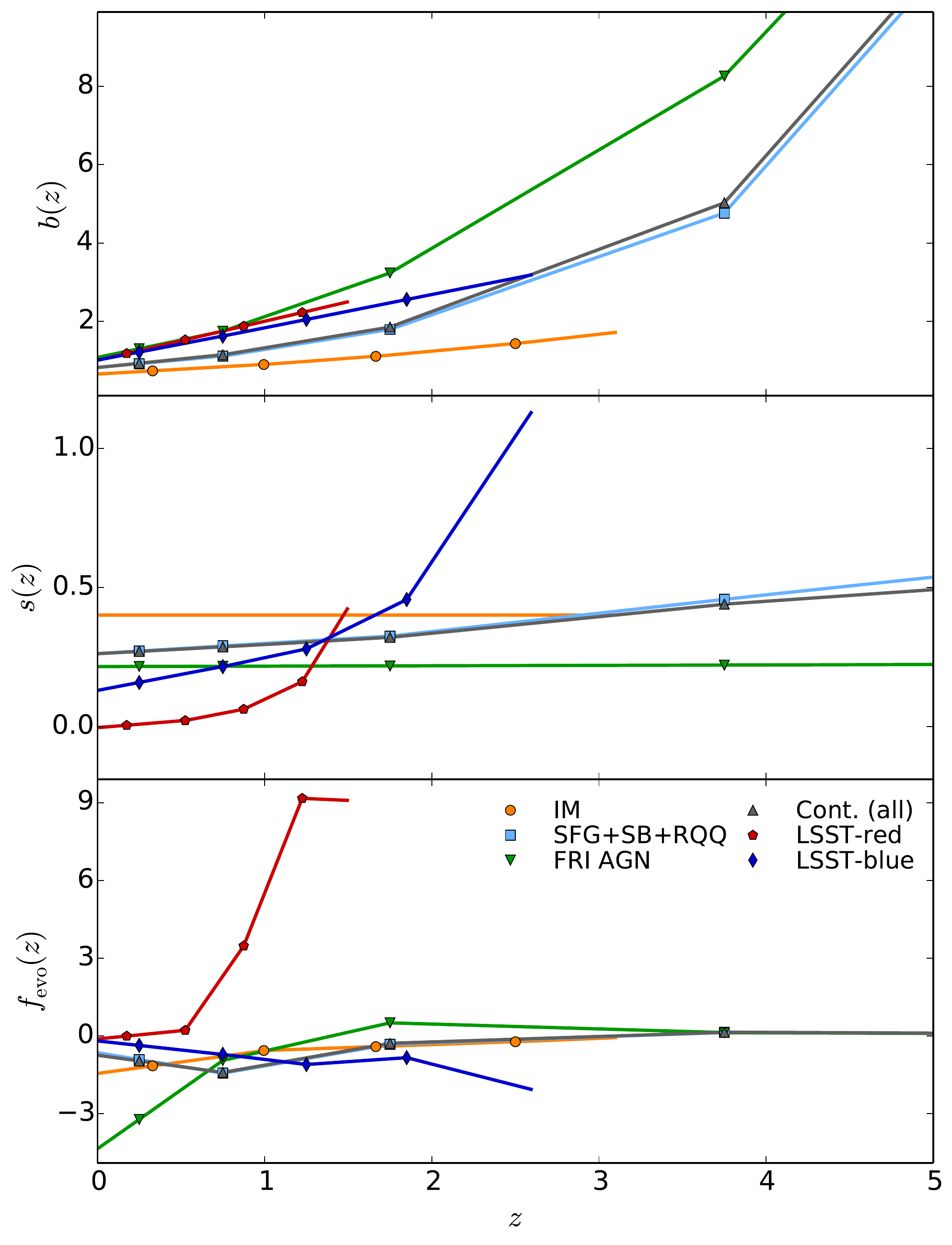}
        \caption{Clustering bias (upper panel), magnification bias (middle panel)
                 and evolution bias (lower panel) for the different tracers
                 considered in this work: intensity mapping (orange),
                 first continuum sample, made up of star-forming galaxies,
                 starbursts and radio-quiet quasars (cyan), second
                 continuum sample, made up of FRI-type AGN (green), combined
                 continuum sample (grey), red LSST galaxies (red)
                 and blue LSST galaxies (blue). In each curve, the markers show
                 the nodes used as free parameters in the marginalization over these
                 bias functions. Notice the very different evolution biases for red
                 and blue galaxies.}
       \label{fig:bias_functions}
      \end{center}
    \end{figure}
    \begin{table}[b]
      \centering{
        \begin{tabular}{|l|l|l|}
          \hline
              {\sf Tracer} & {\sf \# bins} & {\sf Bin edges} \\
              \hline
              Int. map. (SKA1)    & 4 & 0.0, 0.7,  1.3, 2.0,  3.0 \\
              Cont. survey (SKA1) & 4 & 0.0, 0.5,  1.0, 2.5,  5.0\\
              Photo. (LSST), red  & 4 & 0.0, 0.35, 0.7, 1.05, 1.4\\
              Photo. (LSST), blue & 5 & 0.0, 0.5,  1.0, 1.5,  2.2, 3.0\\
              Photo. (DES), red   & 4 & 0.0, 0.3,  0.6, 0.9,  1.2\\
              Photo. (DES), blue  & 4 & 0.0, 0.4,  0.8, 1.2,  1.5\\
              \hline
      \end{tabular}}
      \caption{Redshift bins used for the bias function nuisance parameters,
        for each tracer.}
      \label{tab:nuisancebins}
    \end{table}
    The uncertainty with which a given parameter can be constrained usually
    depends on the largest and smallest accessible scales. The smallest angular
    scale included in our forecasts is given by $\ell_{\rm max}$ in Eq.
    \ref{eq:fisher}, while the smallest radial scale is given by the width of
    the redshift bins used in the analysis. These scales should be chosen on
    the basis of the angular and radial resolution of the experiment
    (e.g. the beam size for intensity mapping or the redshift uncertainty
    for photometric surveys), or in terms of the smallest scales for which we
    can trust our theoretical prediction (e.g. the non-linear scale). Fortunately,
    since the observables we are interested in are mostly relevant on ultra-large
    scales, our forecasts are relatively insensitive to the choice of minimum
    scales. By default we use $\ell_{\rm max}=500$ for all the experiments
    considered here, and address the redshift binning scheme on an individual
    basis. We further assume that all these experiments will be able to probe
    angular scales down to $\ell_{\rm min}=2$, and we asess how the constraints
    degrade as a function of $\ell_{\rm min}$.

    We treat the uncertainty in the bias functions ($b(z),\,s(z)$ and
    $f_{\rm evo}(z)$) using the same method that was used in
    \cite{2015arXiv150507596A}: the mean value for each function is computed in
    a small set of wide redshift bins, and we use the linear interpolation between
    these values as the fiducial bias function passed to {\tt CLASS}. We then treat
    each of the aforementioned mean values as free parameters that we include in
    our computation of the Fisher matrix and marginalize over. The redshift bins
    used to define these bias parameters for the different probes considered here
    are given in Table \ref{tab:nuisancebins}, and the resulting bias functions are
    shown as a function of redshift in Figure \ref{fig:bias_functions}.
    
  \subsection{Why  multi-tracer surveys beat cosmic variance}\label{ssec:multi_why}
    It was first noted in \cite{2009PhRvL.102b1302S} that using different tracers
    deterministically related to the same density field it is possible to measure
    certain parameters with uncertainties smaller than the cosmic-variance limit. The aim
    of this section is to briefly describe the mechanism by which this is possible,
    and how we could benefit from it in order to detect the relativistic corrections
    presented in Section \ref{ssec:th_rel}. 

    Let us consider a simplified scenario, where we measure the fluctuations in the
    source number counts of two different tracers ($\alpha=1,2$) in the same single
    redshift bin. For the sake of the argument let us assume that only two terms
    contribute to the total perturbation:
    \begin{equation}
      a^\alpha_{\ell m}=b_\alpha\,\Delta_{\ell m}^\delta+
      f_\alpha\,\epsilon\,\Delta_{\ell m}^G+n^\alpha_{\ell m}.
    \end{equation}
    Here $\Delta^\delta$ and $\Delta^G$ are the terms corresponding the intrinsic
    perturbation in the matter density and a generic relativistic correction
    respectively, $b_\alpha$ is the clustering bias, $f_\alpha$ is a tracer-dependent
    bias function affecting the GR correction (i.e. similar to the evolution or
    magnification biases) and $n^\alpha_{\ell m}$ is a noise term. We have explicitly
    extracted the amplitude of the GR term as a parameter, $\epsilon$,  equivalent to
    $\egr$. This can be rewritten as
    \begin{equation}\label{eq:multitracer_toy}
      a^\alpha_{\ell m}=\left(b_\alpha\,T^\delta_\ell+
      f_\alpha\,\epsilon\,T_\ell^G\right)\Phi^0_{\ell m}+n^\alpha_{\ell m},
    \end{equation}
    where $T^\delta_\ell$ and $T^G_\ell$ are the transfer functions corresponding to
    the density and GR correction terms, and $\Phi^0_{\ell m}$ is the primordial 
    perturbation.

    The covariance matrix for both tracers in each multipole is
    \begin{widetext}
      \begin{equation}
        \mathsf{C}\equiv 
        \left(\begin{array}{cc}
          b_1^2\mathcal{P}(T^\delta+\beta_1\epsilon T^G)^2+N_1 &
          b_1^2\mathcal{P}(T^\delta+\beta_1\epsilon T^G)(\gamma T^\delta+
          \beta_2\epsilon T^G) \\
          b_1^2\mathcal{P}(T^\delta+\beta_1\epsilon T^G)(\gamma T^\delta+
          \beta_2\epsilon T^G) &
          b_1^2\mathcal{P}(\gamma T^\delta+\beta_2\epsilon T^G)^2+N_2
        \end{array}\right),
      \end{equation}
    \end{widetext}
    where $\mathcal{P}$ is the power spectrum of $\Phi^0$, $\gamma\equiv b_2/b_1$,
    $\beta_1\equiv f_1/b_1$, $\beta_2\equiv f_2/b_1$, $N_\alpha$ is the noise power
    spectrum for each tracer and we have assumed that $\langle n_1 n_2\rangle=0$. Note
    that we have omitted the dependence on the multipole order $\ell$.
    
    We can estimate the best-case uncertainty on the amplitude $\epsilon$ by computing the
    inverse of its Fisher matrix element $\sigma^2(\epsilon)\simeq1/F_{\epsilon\epsilon}$.
    Using $F_{\epsilon\epsilon}=\frac{1}{2}
    \Tr\left[(\mathsf{C}^{-1}\partial_\epsilon\mathsf{C})^2\right]$ we obtain
    \begin{equation}\label{eq:multitracer_error}
     \sigma^2(\epsilon)\simeq
     X_2\,\left[\frac{R+\beta_1\epsilon}{\beta_2-\gamma\beta_1}\right]^2+
     X_1\,\left[\frac{\gamma R+\beta_2\epsilon}{\beta_2-\gamma\beta_1}\right]^2,
    \end{equation}
    where we have defined the ratios $X_\alpha\equiv N_\alpha/(\mathcal{P}\,(b_1 T^\delta)^2)$
    and $R\equiv T^\delta/T^G$, and we have linearized the result with respect to $X_i$.
    Several conclusions can be drawn from Eq. \ref{eq:multitracer_error}:
    \begin{enumerate}
      \item The error on $\epsilon$ is completely limited by the noise level, and not
            by cosmic variance. This could have been anticipated by looking at Equation
            \ref{eq:multitracer_toy} in the noiseless case ($n^\alpha\rightarrow0$):
            by taking the ratio $a^1_{\ell m}/a^2_{\ell m}$ it would be possible to
            cancel the stochasticity of $\Phi^0$ and determine $b_\alpha$, $f_\alpha$
	    and $\epsilon$ free of cosmic variance. This is the key to understanding
	    the multitracer technique.
      \item If the amplitude of the relativistic correction is subdominant,
            (i.e. $R\gg1$), as is the case for the terms studied here, the uncertainty
            in $\epsilon$ becomes proportional to $R$, and thus achieving
            $\sigma(\epsilon)\ll1$ would still require very low noise levels and/or
            appropriate values for the bias parameters $f_\alpha$.
      \item The effect of the factor $\beta_2-\gamma\beta_1$ in the denominator
            is twofold. Firstly, by choosing two tracers with very different bias
            functions (e.g. $f_2\gg f_1$) it is possible to further reduce the
            uncertainty on $\epsilon$. This is key to understand the results presented
            in subsequent sections. Secondly, in the case $\beta_1=\beta_2$ and
            $\gamma=1$ the uncertainty on $\epsilon$ diverges, which is a consequence
            of the fact that in this case both tracers are actually the same one, 
            and nothing can be gained from treating them as being different.
    \end{enumerate}
    
    Although this result is based on a  toy model, it allows us to
    build a picture regarding the optimal tracers that should be used in order to
    achieve a measurement of the GR corrections: the ideal case would be to
    combine measurements of two (or more) high signal-to-noise tracers with very different
    magnification and/or evolution biases.

\section{Optical and near-infrared experiments}\label{sec:photo}
  \subsection{Photometric redshift surveys}\label{ssec:photo_observables}
    The observational study of cosmological large-scale structure has so far been
    dominated by galaxy surveys performed in optical and infrared wavelengths
    \cite{2011MNRAS.416.3017B,2011MNRAS.415.2892B,SDSS.DR9}. In the ideal case,
    redshifts and angular positions are measured for a homogeneous sample of targets,
    and the three-dimensional matter density distribution is probed through the
    fluctuations in the galaxy number density. Accurate redshift measurements are, however,
    very time consuming, especially for faint sources, and therefore spectroscopic redshifts
    are usually only available for a subsample of all the photometrically-detected sources.
    This has two critical drawbacks: firstly, it prevents spectroscopic redshift surveys
    from covering the largest accessible scales, which is key to studying the observables
    described in section \ref{sec:th_uls}. Secondly, the sub-sampling of the galaxy
    population produces relatively high shot-noise levels, which dominate the
    error budget for parameters like $\fnl$ and $\egr$ in the multi-tracer scheme. These
    problems can be partially circumvented using photometric redshifts (photo-$z$s), i.e.
    redshifts inferred from the observed source fluxes in a number of broad bands. On
    the one hand, these redshift estimates are much less accurate than their spectroscopic
    counterparts, which limits the ability of photometric redshift surveys to study
    clustering statistics on small radial scales. On the other hand, photo-$z$s can, in
    principle, be estimated for all the sources detected above a given flux threshold,
    and thus the problems of volume coverage and shot noise are greatly alleviated.
    Since the observables studied  here are only relevant on the largest scales, and
    their measurement can be optimized by reducing the noise levels of the experiment,
    photometric redshift surveys seem to be ideally suited for our purposes.

    We will focus our discussion on the Large Synoptic Survey Telescope 
    (LSST, \cite{2009arXiv0912.0201L}), since it represents the widest
    ($\sim20000\,{\rm deg}^2$) and deepest ($r\sim27$) photometric survey planned in the
    foreseeable future, but we will also present forecasts for the  ongoing Dark Energy
    Survey \cite{2005astro.ph.10346T}.
    
    In this section we focus on the advantages of combining clustering information for
    different galaxy samples observed within the same experiment. In particular we will
    use a simplified picture in which we assume that the full galaxy population can be
    separated into two disjoint samples of ``red'' and ``blue'' galaxies:
    \begin{itemize}
      \item Red galaxies are usually identified with an early population of galaxies where
            star formation has ended. This produces a red spectrum, usually characterized
            by a deep Balmer/4000\AA{} break. This spectral type is also commonly
            associated with an elliptical morphology, supported by random motions rather
            than angular momentum, possibly caused by mergers. The red population appears
            only after the peak in the star formation rate ($z\sim1.9$), and its comoving
            number density is known to decrease sharply at redshifts larger than $\sim 1$. 
            
            The simplest models of galaxy formation suggest that the earliest galaxies
            should have formed in the peaks of the density field. Thus the red population
            can be associated with larger-mass halos and a larger clustering bias.
      \item Blue galaxies, on the other hand, are characterized by a population of young
            stars, showing strong Balmer emission lines. They are usually associated with
            a spiral morphology, smaller halo mass and lower clustering bias. The number
            density of blue galaxies shows no sharp increase or decrease, and LSST should
            be able to observe them up to much higher redshifts than the red population.
    \end{itemize}
    This simplified model of the galaxy population is based on the observed bimodal
    nature of the color-magnitude diagram \cite{2001AJ....122.1861S}. The actual picture
    is nowhere as simple (e.g. red spirals and blue ellipticals are known to exist), but
    this model allows us to study the potential improvement in the figure of merit for
    ultra large-scale observables caused by combining clustering observations for two
    populations with similar characteristics.
    
    \begin{figure*}
      \begin{center}
        \includegraphics[width=0.49\textwidth]{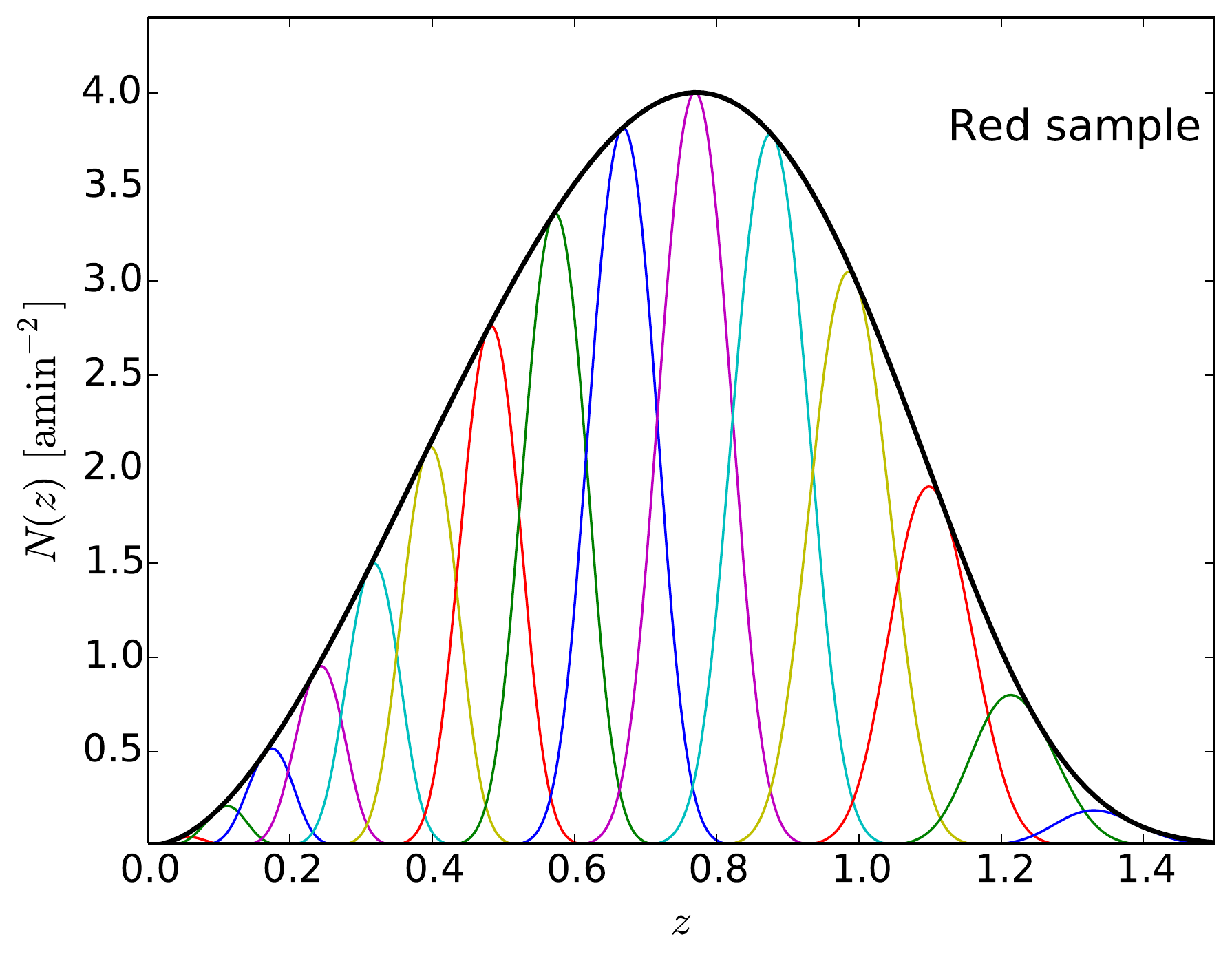}
        \includegraphics[width=0.49\textwidth]{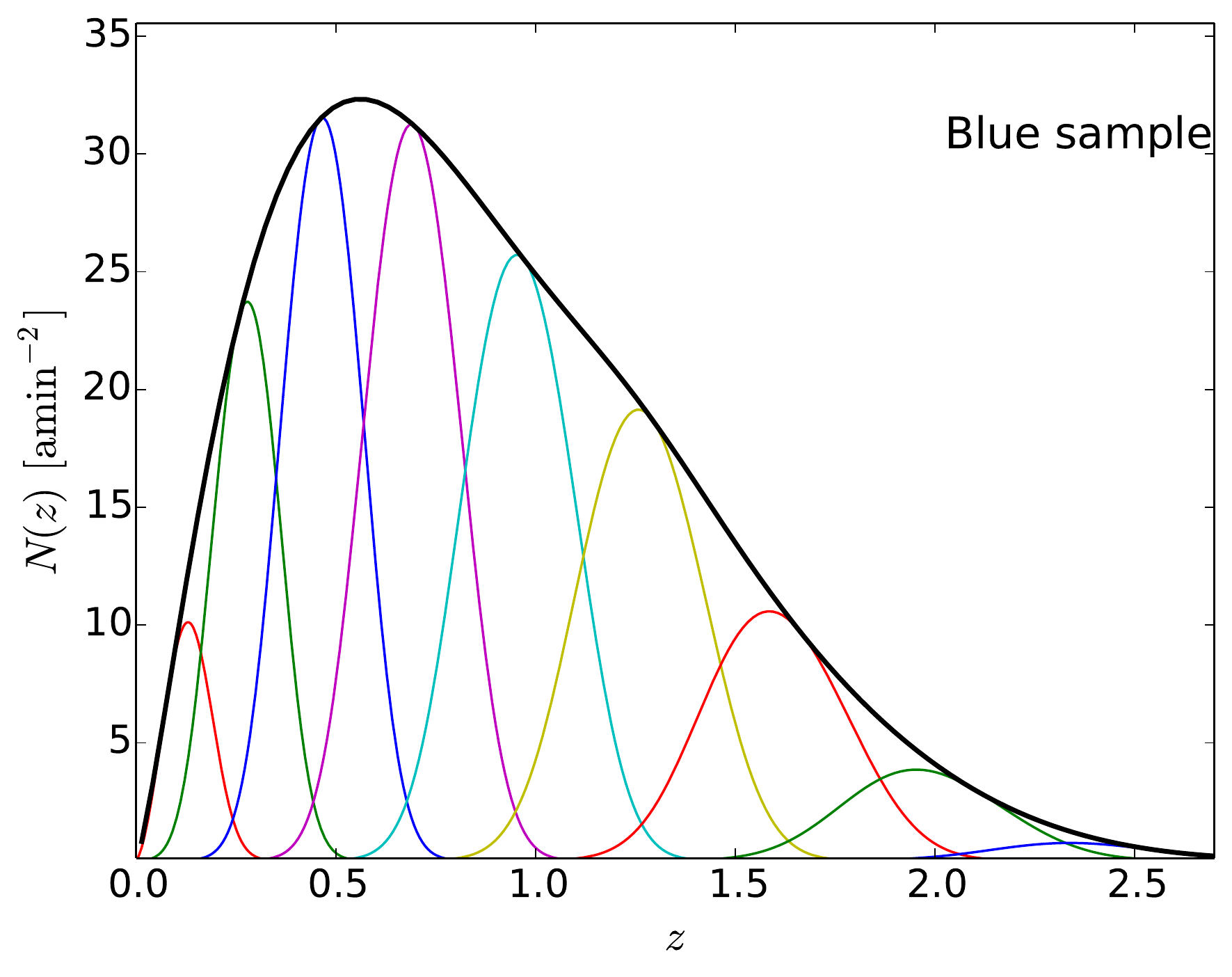}
        \caption{Redshift distribution (thick solid black lines)
                 and radial window functions (thin coloured lines) for the
                 redshift bins used here for the red (left panel) and blue
                 (right panel) galaxy populations for LSST.}
       \label{fig:nz_lsst}
      \end{center}
    \end{figure*}
    Of particular interest would be the sharp decay in the number density of red
    galaxies. As was shown in \cite{2015arXiv150507596A}, this is associated with a
    large evolution bias, and enhances the amplitude of the relativistic corrections
    described in section \ref{ssec:th_rel}, although this enhancement was shown to be
    too small to grant a detection of these effects using single-tracer observations.
    However, as discussed in section \ref{ssec:multi_why}, combining red and blue
    galaxies, the latter having a much lower evolution bias, could potentially boost
    the figure of merit for $\egr$.
    
    As was done in \cite{2015arXiv150507596A}, we base our estimates for the number
    densities of both populations, as well as their magnification and evolution biases
    on luminosity functions estimated by \cite{2007ApJ...665..265F,2006A&A...448..101G}
    in the $r$-band. The details of this model are summarized in \cite{2015arXiv150507596A}.
    Regarding the clustering bias, we parametrize its redshift dependence for the red and
    full galaxy samples as:
    \begin{equation}
      b_{\rm red}(z)=1+z,\hspace{12pt} b_{\rm full}(z)=1+0.84\,z,
    \end{equation}
    based on the simulations of \cite{2004ApJ...601....1W} and the measurements of
    \cite{2008ApJ...672..153C}. We then assume that the clustering bias of the full
    sample is a weighted average of the red and blue samples, and compute the
    clustering bias of the latter as:
    \begin{equation}
      b_{\rm blue}(z)=\frac{\bar{n}_{\rm full}(z)b_{\rm full}(z)-
                            \bar{n}_{\rm red}(z)b_{\rm red}(z)}{\bar{n}_{\rm blue}(z)}.
    \end{equation}    
    
    Finally we assume a magnitude cut of $i=25.3$ ($r\simeq26$), corresponding to the
    so-called ``gold sample'', and a sky fraction $f_{\rm sky}=0.5$ ($\sim\,20000\,
    {\rm deg}^2$) for LSST. We find that the total number counts as well as the magnitude
    distribution for each galaxy type obtained using these models is in good agreement
    with previous estimates \cite{2006A&A...457..841I,2009arXiv0912.0201L}. For the Dark
    Energy Survey we will assume a magnitude limit $r=24$ and a sky fraction of $0.125$
    ($\sim5000\,{\rm deg}^2$).
    
    The redshift binning scheme was determined for both samples in terms of their expected
    photometric redshift errors. We assume Gaussianly distributed errors with a variance
    $\sigma_z(z)=\sigma_0(1+z)$. The photo-$z$ requirements for LSST as quoted in 
    \cite{2009arXiv0912.0201L} are $\sigma_0<0.05$ with a goal of $0.02$. Since the
    spectral characteristics of red galaxies, particularly the prominent Balmer break,
    make their photo-$z$s more reliable, we assumed that the goal quoted above will be
    attained for this sample, and we conservatively set the photo-$z$ error parameter for
    blue galaxies to the requirement limit. I.e.:
    \begin{equation}
      \sigma_0^{\rm red}=0.02,\hspace{12pt}\sigma_0^{\rm blue}=0.05.
    \end{equation}
    For DES we kept $\sigma_0^{\rm blue}=0.05$ but increased the error parameter for
    red galaxies to $\sigma_0^{\rm red}=0.03$, since the quality of the photo-$z$s should
    be degraded by the absence of the $u$-band in the DES filter set. We define the
    redshift bins used for each sample as top-hat bins in the space of photo-$z$s with a
    width three times the size of the photo-$z$ uncertainty at the median redshift of the
    bin. This width was chosen in order to reduce the correlations between redshift bins
    caused by the tails of the photo-$z$ probability distribution. Thus, a top-hat bin
    in photo-$z$ space with edges $z_0$ and $z_f$ has a true-$z$ window function given by
    \begin{equation}
      W(z)\propto \bar{N}(z)\,w(z|z_0,z_f),
    \end{equation}
    where $w(z|z_0,z_f)$ is the integral of the photo-$z$ distribution over the
    redshift bin. For the Gaussian photo-$z$ errors assumed here, $w$ takes the form
    \begin{equation}\label{eq:win_zp}
      w(z|z_0,z_f)=\frac{1}{2}\left[{\rm erf}\left(\frac{z-z_0}{\sqrt{2}\sigma_z}\right)
                                   -{\rm erf}\left(\frac{z-z_f}{\sqrt{2}\sigma_z}\right)
      \right].
    \end{equation}
    Using this prescription we divide the red and blue samples into 15 and 9 redshift bins
    for LSST (10 and 7 bins respectively for DES), shown in figure \ref{fig:nz_lsst}
    together with their overall redshift distribution.
    
    The noise power-spectrum is given by $N_\ell^{ij}=\delta_{ij}/n^i$, where
    \begin{equation}
      n^i\equiv\int_0^\infty dz\,\bar{N}(z)w^i(z)
    \end{equation}
    is the angular number density of sources in the $i$-th redshift bin with a photo-$z$
    window function $w^i$, given by Eq. \ref{eq:win_zp}.
    \begin{figure}
      \begin{center}
        \includegraphics[width=0.49\textwidth]{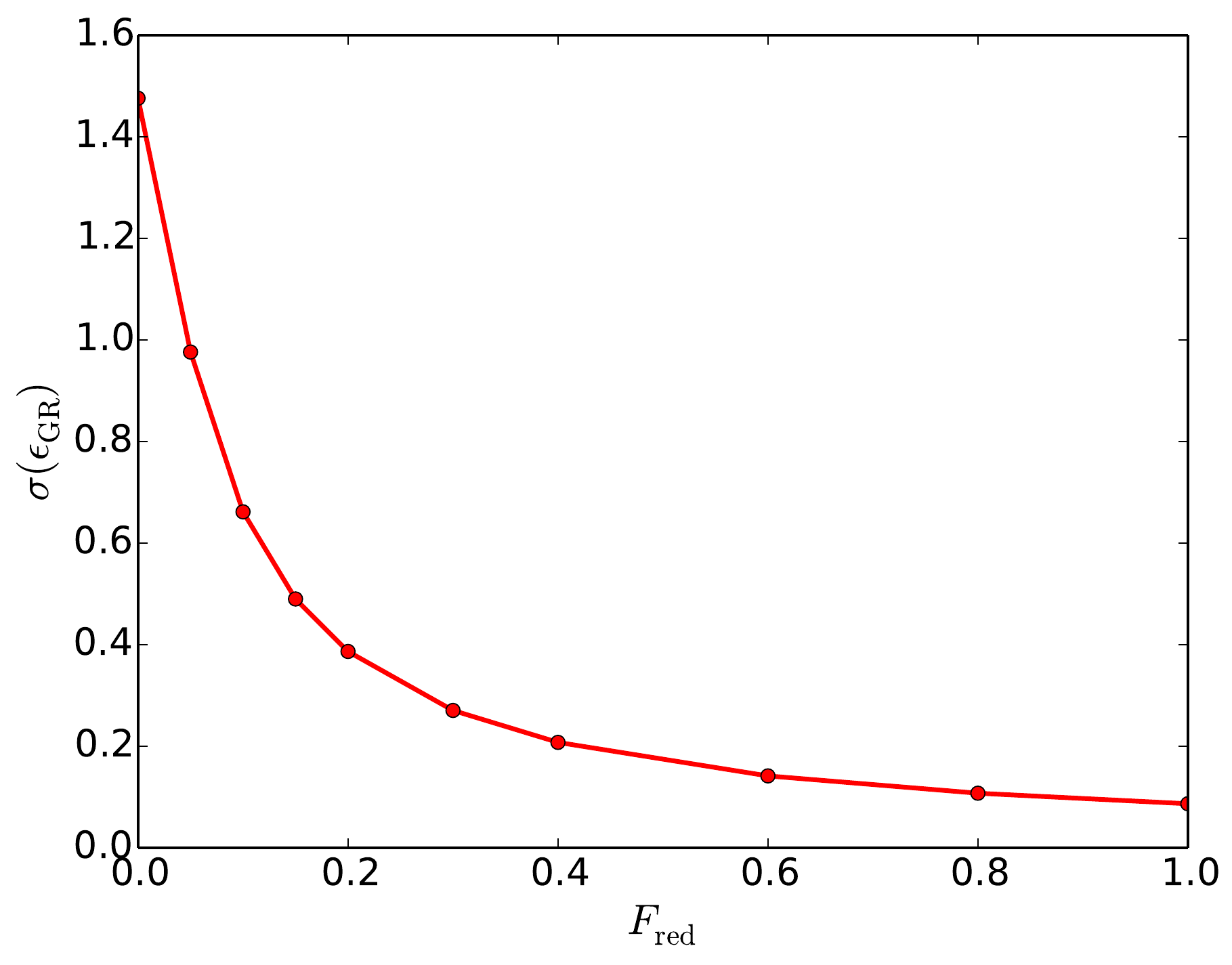}
        \caption{Improvement of the detection level for the relativistic corrections
                 as a function of $F_{\rm red}$ (see Eq. \ref{eq:fred}), which
                 parametrizes the amplitude of the difference in the values of $\fevo$
                 and $s$ between red and blue galaxies. As discussed in Section
                 \ref{ssec:multi_why} the uncertainty on $\egr$ decreases as the
                 values of these two parameters for both populations become more
                 different.}
       \label{fig:segr_fred}
      \end{center}
    \end{figure}
    \begin{figure*}
      \begin{center}
        \includegraphics[width=0.49\textwidth]{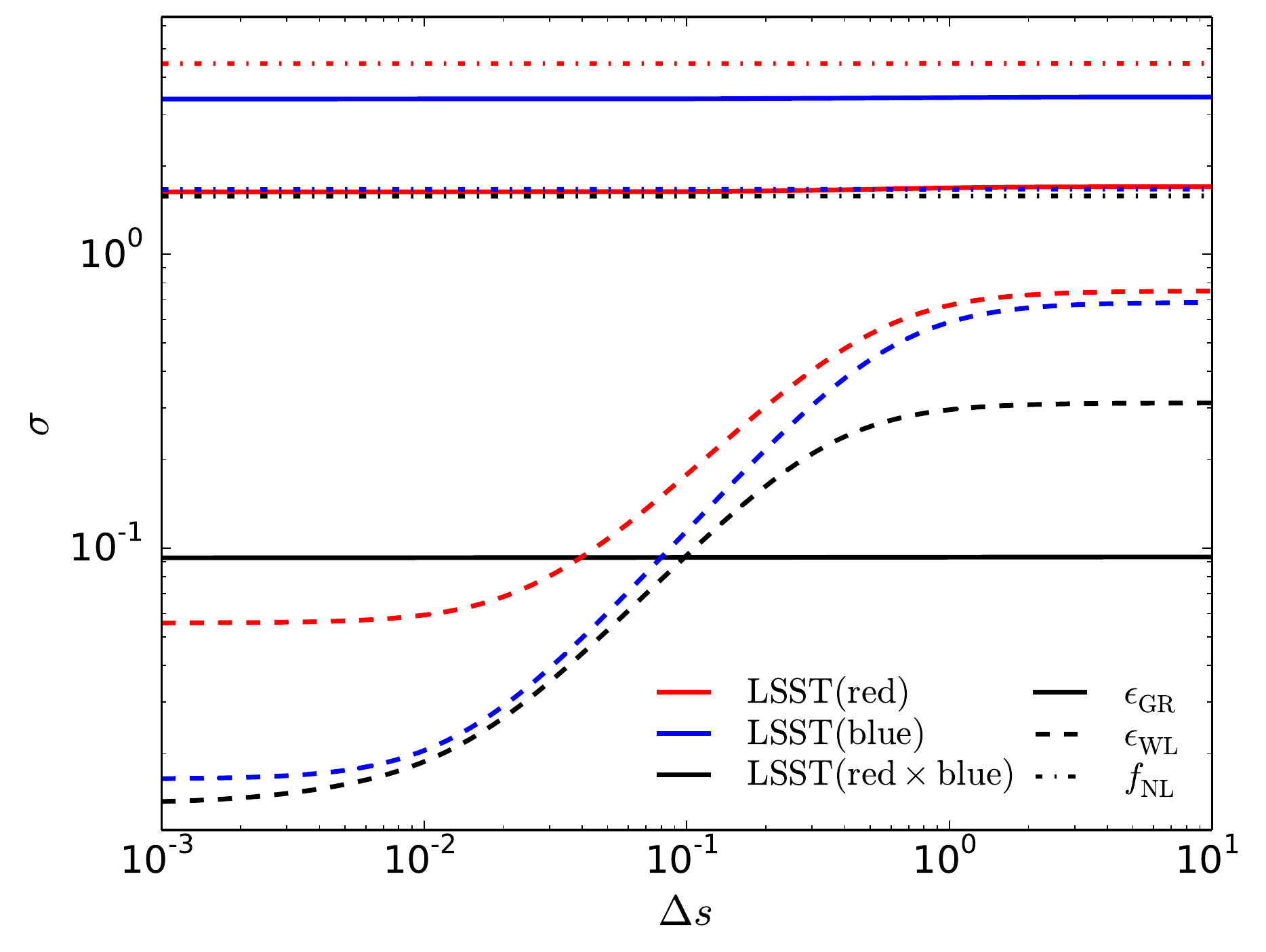}
        \includegraphics[width=0.49\textwidth]{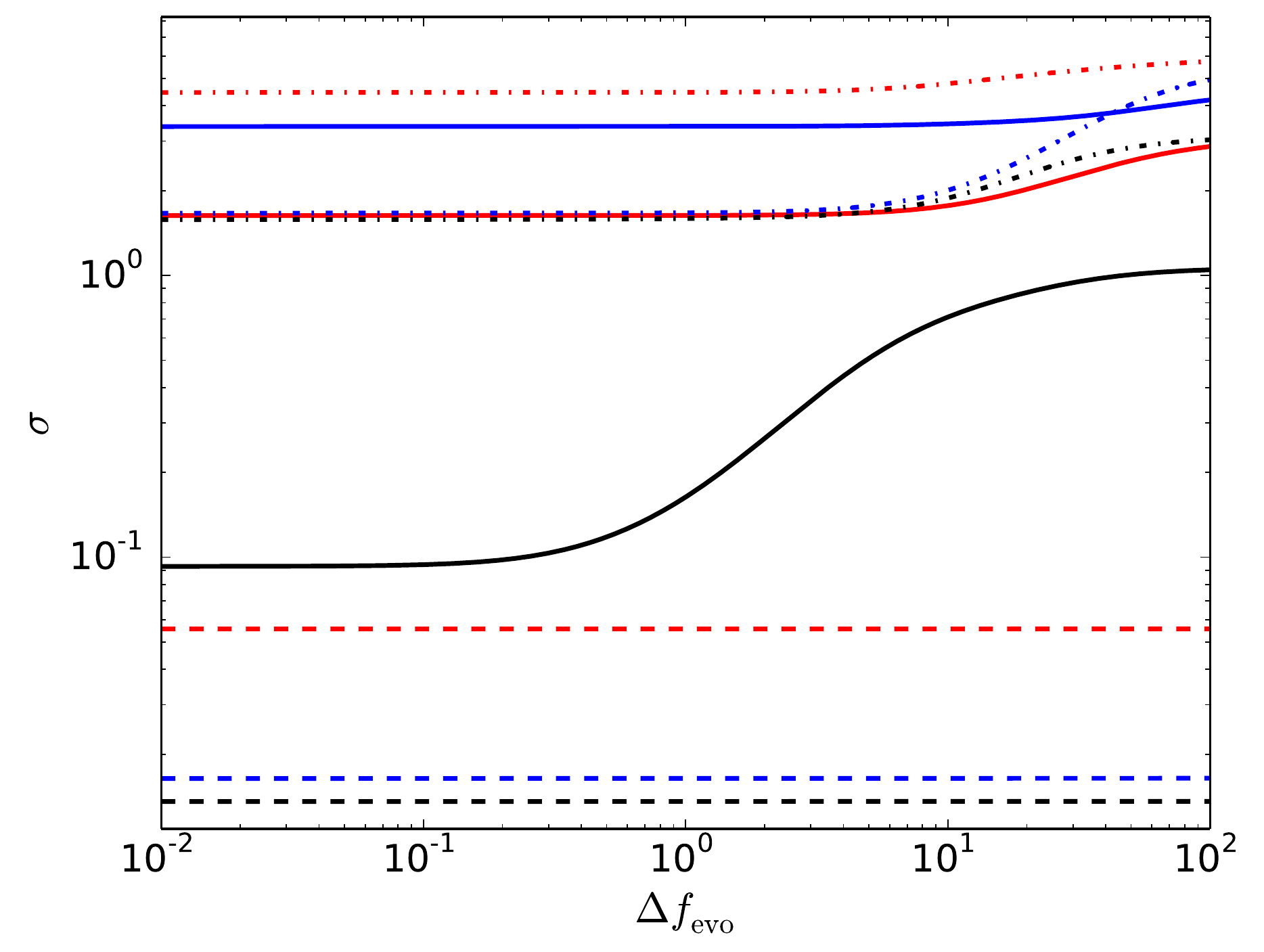}
        \caption{Dependence of the uncertainty on $\fnl$, $\egr$ and $\ewl$ on the
                 assumed Gaussian constant prior on the magnification bias (left panel) and
                 the evolution bias (right panel). The significance of the detection of
                 the lensing magnification term depends very strongly on the prior on $s$.
                 Likewise the error on $\egr$ is particularly sensitive to the prior on
                 $\fevo$ when both tracers are combined, and the detection of GR effects
                 only becomes optimal for $\Delta\fevo\lesssim0.1$.}
       \label{fig:sigmas_nuisance_lsst}
      \end{center}
    \end{figure*}

  \subsection{Forecasts}\label{ssec:photo_forecasts}
    \begin{figure}
      \begin{center}
        \includegraphics[width=0.49\textwidth]{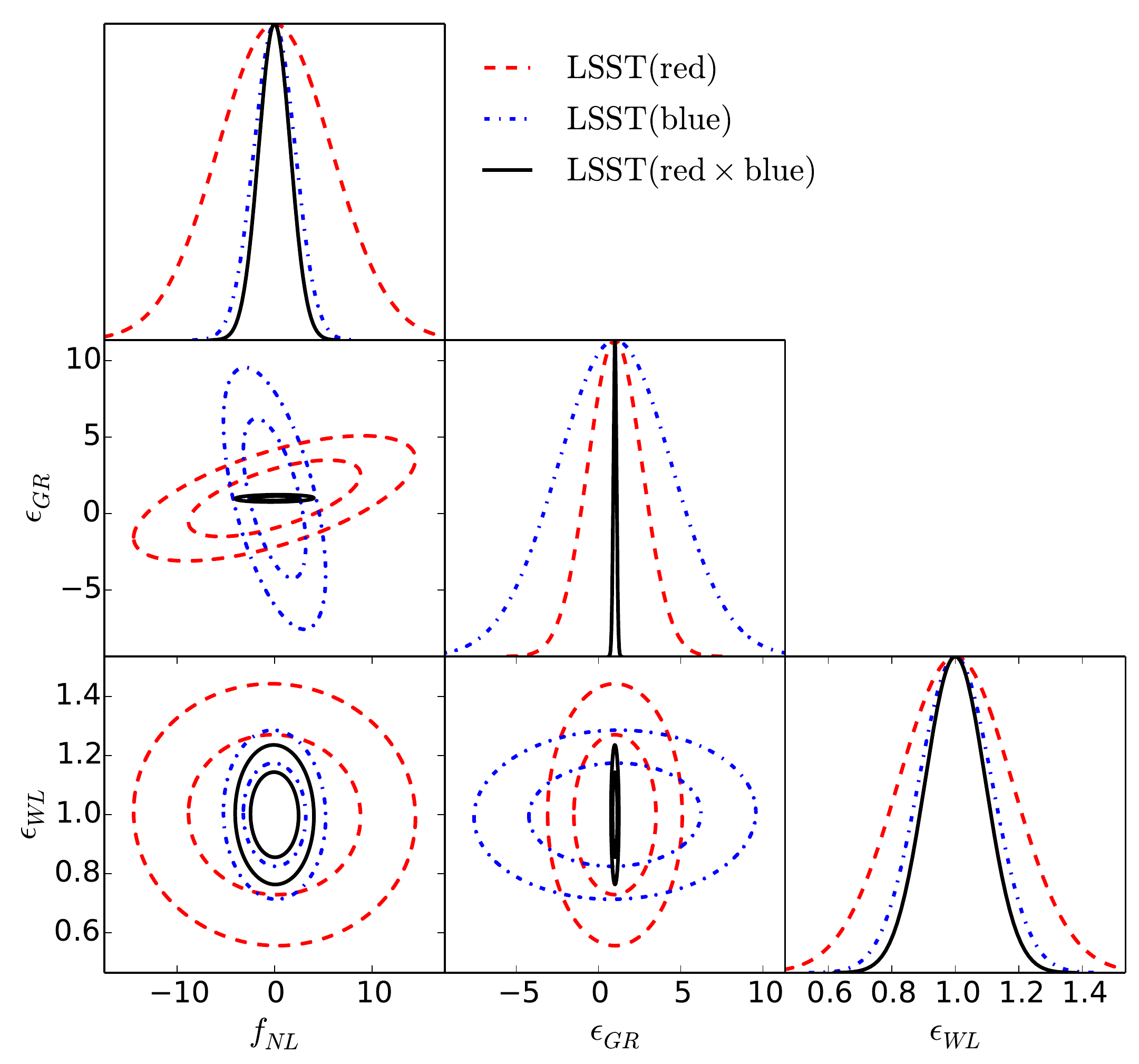}
        \caption{Joint constraints on $\fnl$, $\egr$ and $\ewl$ for the red
                 and blue samples (red and blue contours respectively) as well
                 as for a joint analysis of both (black contours) for optimistic
                 priors $\Delta\fevo=\Delta s=0.1$.}
       \label{fig:ellipses_lsst}
      \end{center}
    \end{figure}
    We begin our discussion with an assessment of the detectability of the GR
    effects in the simplest and most optimistic case: we marginalize only over
    $\egr$ and $\fnl$ (since the latter could potentially be very degenerate with
    the former) and fix all other parameters to their fiducial values. This would
    represent what an experiment aiming at a first detection of the GR corrections
    would do. The results using the red sample only, the blue sample only and
    the combination of both (in a multi-tracer sense) are:
    \begin{align}\nonumber
      \sigma_{\rm red}(\egr)=1.531,\hspace{6pt}
      &\sigma_{\rm blue}(\egr)=3.094,\\\nonumber
      \sigma_{\rm joint}(\egr)&=0.095.
    \end{align}
    We can see that the combination of both samples causes a dramatic improvement
    in the figure of merit. From our discussion in section \ref{ssec:multi_why}, we
    can a priori attribute this to two causes: the low shot-noise levels for LSST
    and the very different bias parameters of blue and red galaxies, especially
    the evolution bias, as can be seen in Fig. \ref{fig:bias_functions}.
    
    In order to show this explicitly we have carried out the following exercise.
    We recomputed our forecasts for a fictitious LSST experiment with an ordinary
    blue sample and a red sample with the same number density and clustering bias
    shown in Figures \ref{fig:nz_lsst} and \ref{fig:bias_functions}, but with
    magnification and evolution biases given in terms of a parameter $F_{\rm red}$ as:
    \begin{align}\nonumber
      s^{\rm red}\rightarrow F_{\rm red} s^{\rm red}+(1-F_{\rm red})s^{\rm blue},\\
      \label{eq:fred}
      \fevo^{\rm red}\rightarrow
      F_{\rm red} \fevo^{\rm red}+(1-F_{\rm red})\fevo^{\rm blue}.
    \end{align}
    For a value of $F_{\rm red}=0$ both samples would have the same magnification
    and evolution biases and, according to our discussion in Section \ref{ssec:multi_why}
    the uncertainty on $\egr$ would not suffer any significant enhancement. As
    we increase the value of $F_{\rm red}$ towards $1$ we should gradually recover
    the result quoted above. This is explicitly shown in Figure \ref{fig:segr_fred},
    which presents the forecasted uncertainty on $\egr$ as a function of $F_{\rm red}$.
    A $2\sigma$ detection of $\egr$ would already be achievable, in this optimistic setup,
    just with $F_{\rm red}\sim	0.2$.

    Since it seems like it could be possible to detect the relativistic corrections
    using the multi-tracer method, we must now explore the possible degereracies that
    could, on the one hand, preclude this detection and, on the other, affect the
    measurements of other parameters. We do not observe strong degeneracies between
    $\egr$ and any cosmological parameter, and only a very mild one with $\fnl$. The
    situation is different, however, for the nuisance bias parameters. Since the
    amplitude of the relativistic corrections is partially determined by the values
    of the magnification and evolution biases of the samples under study, it is probably
    very difficult to claim a detection of these effects without good prior information
    on $s$ and $f_{\rm evo}$. We have studied the dependence of the uncertainties on
    $\fnl$, $\egr$ and $\ewl$ on a Gaussian prior, constant in redshift, for these two
    bias parameters. The results are shown in Fig. \ref{fig:sigmas_nuisance_lsst} in
    the same three cases explored before (red and blue samples independently and jointly).
    As should be obvious, the detection level of the lensing magnification term depends
    critically on the precision with which the magnification bias is known, and an error on
    $s(z)$ of the order of $\sim0.01$ would be necessary to optimize this measurement. On
    the other hand both $\fnl$ and $\egr$ are almost insensitive to the prior on $s$. The
    situation is different for $\fevo$: both $\fnl$ and $\egr$ are sensitive to the
    prior on $\fevo$, although only a very broad prior of $\Delta\fevo=\mathcal{O}(1)$
    would be necessary to optimize their measurement in the single-tracer case. For the
    multi-tracer analysis, however, the detection of the GR corrections depends critically
    on having precise enough measurements of the evolution bias, which as we saw is
    responsible for significantly boosting the signal. The optimal case would correspond
    to a precision of $\Delta\fevo\lesssim0.1$. We have also studied the possible dependence
    of the uncertainties on our key observables with the priors on the clustering bias,
    finding no significance degeneracies.
    
    The availability of such tight priors on the evolution and magnification biases
    will depend directly on our ability to model the joint redshift and magnitude
    distribution of the galaxy samples under study. This could be challenging for
    photometric redshift surveys, given the poor redshift uncertainties, and would
    probably entail relying on external datasets or on a spectroscopic follow-up of a
    representative subsample of the survey. In order to gain some intuition regarding
    the feasibility of these measurements we have estimated the uncertainty on the
    evolution bias of red galaxies (responsible for the enhanced amplitude of the GR
    effects) given the uncertainties in the measurements of the luminosity function of
    red galaxies of \cite{2007ApJ...665..265F} on which our model is based. In the
    high-redshift end of the distribution we obtain an uncertainty $\Delta\fevo\sim2$,
    which decreases towards smaller redshifts as the number counts increase. In the
    future these uncertainties should improve significantly due to the availability of
    larger and deeper datasets, and therefore we will present our forecasts in a
    pessimistic and an optimistic scenario, where we will assume the priors:
    \begin{align}\nonumber
      \text{Pessimistic}\rightarrow &\Delta s=\Delta\fevo=1,\\\nonumber
      \text{Optimistic}\rightarrow &\Delta s=\Delta\fevo=0.1.
    \end{align}

    Figure \ref{fig:ellipses_lsst} presents the joint forecasted uncertainties on $\fnl$,
    $\egr$ and $\ewl$ for LSST in the optimistic scenario for the three cases mentioned
    above: red-only (red), blue-only (blue) and joint analysis (black). The numerical
    values for these uncertainties are given in Table \ref{tbl:constraints_summary}.
    Note that, unlike in the case of $\egr$ there is only a mild improvement in the
    figure of merit for $\fnl$ and $\ewl$ in the multi-tracer analysis with respect
    to the deepest LSST blue sample alone. The main reason for this is that, unlike
    in the case of the evolution bias, the differences in the magnification and
    clustering biases of both tracers are not so large.
    
    We have also produced forecasts for the Dark Energy Survey, using the same models
    adopted for LSST with a magnitude limit of $r=24$ and $f_{\rm sky}=1/8$. The results
    are also included in Table \ref{tbl:constraints_summary}: DES should, in the best-case
    scenario, be able to make a $\sim3\sigma$ detection of the relativistic corrections.
    This result, however, could be compromised by the possible systematic effects that
    could dominate the clustering statistics on large angular scales. We will discuss these
    in Section \ref{sec:discussion}.

\section{Radio experiments}\label{sec:radio}
  \subsection{Cosmological radio surveys}\label{ssec:radio_observables}
    With the forthcoming wide-area radioastronomy facilities,
    the field of observational large-scale structure will soon begin to reap the
    benefits of observing in the radio regime. The low atmospheric absorption
    and dust obscuration in a wide range of radio frequencies makes it possible
    to observe objects at significantly higher redshifts than are usually targeted
    in optical/NIR surveys, and in the next decades radio surveys will be able
    to cover comparably wide areas with similar source number densities. In addition
    to that, the relative isolation of the few emission lines of astrophysical interest
    in the radio spectrum (e.g. the neutral hydrogen line at 1.4 GHz or molecular CO at
    115 GHz) makes it possible to conduct intensity mapping observations, producing
    tomographic maps of the density fluctuations of these species.
    
    In this section we will describe two main cosmological probes of the low-redshift
    Universe in radio experiments: intensity mapping (IM) of neutral hydrogen (HI) and continuum
    surveys. Both probes have complementary properties that make their cross-correlation
    extremely interesting: while intensity mapping is able to map the density field
    with remarkable radial (frequency) resolution, but only on fairly big angular scales,
    continuum surveys lack any radial information, but are able to resolve
    individual sources. On the other hand, while HI is known to be a very faithful
    tracer of the matter density (i.e. has a low clustering bias), many radio sources,
    such as AGNs, are highly biased, and thus, as discussed in Section \ref{ssec:multi_why},
    their combination could be a useful way to measure the level of primordial
    non-Gaussianity below the cosmic variance limit.
    
    For both probes we produce our forecasts assuming the current design for Phase 1
    of the Square Kilometre Array (SKA) \cite{2009IEEEP..97.1482D,braun15}. In both
    cases we assume that the experiment of choice will be SKA1-MID, an array of
    $\sim200$ single-pixel, 15m dishes to be installed in South Africa. It will cover
    the frequency range $350-1760$ MHz ($z\lesssim3$) in two separate bands. Note
    that there are other intensity mapping experiments planned for the future, such
    as CHIME \cite{2014SPIE.9145E..22B} or BINGO \cite{2012arXiv1209.1041B}. However,
    the SKA should provide a better coverage of the largest angular scales, which is
    the topic of this work. Additionally, the SKA is a multi-science facility that will
    perform both intensity mapping and continuum observations, and therefore cross
    correlating both datasets from the same experiment should be more straightforward.
    
    \subsubsection{Intensity mapping}\label{sssec:im}
      While it is, in principle, possible to determine the radial position of radio sources
      through the redshift of the 21cm line, the intensity of this line from individual
      sources is very small, and long integration times are needed in order to resolve it
      with sufficiently  high signal-to-noise ratio. For a fixed amount of observation
      time, this limits the source number density, depth and area that can be achieved
      with such surveys. Intensity mapping \cite{Battye:2004re, 2008MNRAS.383..606W,
      Chang:2007xk} is a relatively novel technique that aims to circunvent this problem
      by measuring the combined emission from all the sources in relatively wide pixels
      simultaneously at different frequencies. Adding up the emission from hundreds or
      thousands of sources it is possible to detect the large-scale fluctuations in the HI
      density field in three dimensions up to relatively large depths. This is achieved at
      the cost of losing potentially valuable information on small angular scales, since
      individual sources are not resolved, although this is not necessarily an issue if we
      target large-scale observables, as is the case of this work. As was shown in
      \cite{2015arXiv150507596A}, however, two facts conspire against HI as a single-probe
      of non-Gaussianity and GR corrections: its relatively low clustering bias and the
      cancellation of the linear perturbations on angular distances respectively.
      
      We model the clustering, magnification and evolution biases for HI intensity mapping
      using the same approach described in section 4.2 of \cite{2015arXiv150507596A},
      based on using a scaling relation between HI and halo mass of the form
      $M_{\rm HI}(z,M) \propto M^\alpha$, with $\alpha\simeq0.6$. This allows us to
      estimate the HI density (and hence its evolution bias using equation
      \ref{eq:fevo_im}), as well as its clustering bias. As noted in section
      \ref{ssec:th_rel}, the exact cancellation of the perturbations on transverse
      distances for intensity mapping implies an effective magnification bias
      $s_{\rm HI}=2/5$.
      
      As described in \cite{2015ApJ...803...21B}, the most efficient use of SKA1-MID
      for intensity mapping will be in single-dish mode, and we base our noise model
      on this configuration. For an autocorrelation experiment, and assuming no
      correlations between the noise in different frequency channels, the noise power
      spectrum is constant on all scales:
      \begin{equation}
        N^{{\rm HI},ij}_\ell=\delta_{ij}\sigma_i^2,
      \end{equation}
      where $\sigma^2_i$ is the noise variance per steradian in the $i$-th frequency
      channel, which can be related to the survey characteristics as
      \begin{equation}
        \sigma_i^2=\frac{T_{\rm sys}^2(\nu_i)\,4\pi\,f_{\rm sky}}
        {\delta\nu\,t_{\rm tot}\,N_{\rm dish}},
      \end{equation}
      where $t_{\rm tot}$ is the total observation time, $f_{\rm sky}$ is the observed
      sky fraction, $N_{\rm dish}$ is the number of dishes in the experiment, $\delta\nu$
      is the frequency bandwidth of the bin and $T_{\rm sys}=T_{\rm sky}+T_{\rm inst}$ is
      the system temperature, which receives contributions both from the instrumental noise
      temperature and from the atmospheric and background noise ($T_{\rm sky}\sim60\,
      {\rm K}\times(\nu/300\,{\rm MHz})^{-2.5}$).

      Regarding the radial binning, we follow the procedure used in
      \cite{2015arXiv150507596A}, subdividing the redshift range ($0\leq z \leq 3$)
      into intervals of equal comoving width. We produced forecasts for gradually thinner
      bins until the uncertainties on our large-scale parameters,  $\fnl$ and $\egr$
      converged. We thus obtain our final forecasts for 100 frequency bins with a
      comoving width $\delta\chi\simeq44\,{\rm Mpc}/h$.
      
      Since, in intensity mapping, sources are not detected individually, we must incorporate
      the effect of the telescope beam, and thus our model for the intensity mapping power
      spectrum is
      \begin{equation}
        C_\ell^{ij}=C_\ell^{S,ij} B^{i}_\ell B^{j}_\ell+N^{ij}_\ell,
      \end{equation}
      where $B^{i}_\ell$ is the harmonic transform of the instrumental beam in the $i-$th
      frequency bin. For this work we have assumed that the beams are isotropic and Gaussian:
      $B^i_\ell=\exp(-\ell(\ell+1)\,\theta_B^2/2)$, with
      $\theta_B(\nu)=c/(2.35\,\nu\,D_{\rm dish})$, where $D_{\rm dish}$ is the dish
      diameter. Note that when cross-correlating intensity mapping with any discrete tracer
      $\alpha$, the effect of the beam is:
      \begin{equation}
        C_\ell^{({\rm HI},i)(\alpha,j)}=B^i_\ell\,C^{S,({\rm HI},i)(\alpha,j)}_\ell.
      \end{equation}
      
      The experiment parameters used for SKA1-MID were: $T_{\rm inst}=25\,{\rm K}$,
      $f_{\rm sky}=0.75$ ($\sim30000\,{\rm deg}^2$), $t_{\rm tot}=10^4\,{\rm h}$,
      $N_{\rm dish}=200$ and $D_{\rm dish}=15\,{\rm m}$.

    \subsubsection{Continuum surveys}\label{sssec:continuum}
      \begin{figure}
        \begin{center}
          \includegraphics[width=0.49\textwidth]{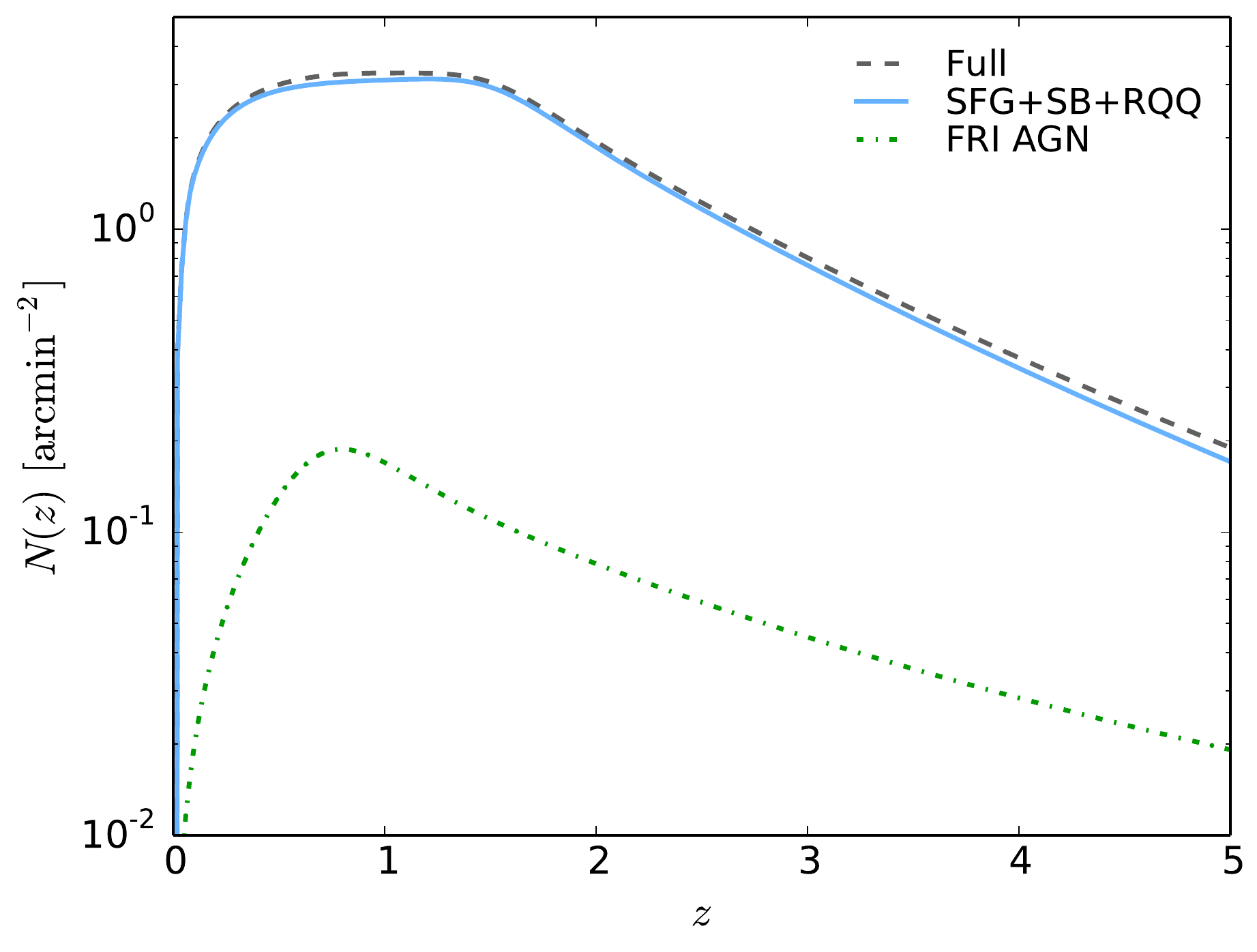}
          \caption{Redshift distribution of radio sources for a flux cut $S<5\uJy$. The
                   distribution for the combined population is shown in dark grey,
                   while the two subsamples considered here are plotted in blue (sample 1 - 
                   star-forming galaxies, starbursts and radio-quiet AGN) and green 
                   (sample 2 - FRI AGN).}
         \label{fig:nz_continuum}
        \end{center}
      \end{figure}
      The radio spectrum of most astrophysical sources is generally smooth and featurless,
      dominated by radio synchrotron emission with, perhaps, very few emission lines
      such as the aforementioned 21cm signal. It is thus possible to integrate the flux
      from individual sources on rather wide frequency bands without losing too much
      information, while gaining access to much fainter sources than it would be possible
      to detect otherwise. Continuum surveys can therefore observe radio galaxies over
      incredibly large volumes with the caveat that all information regarding the radial
      distribution and clustering of these sources is completely unavailable.
      
      A number of papers have already shown the power of continuum surveys to constrain
      the level of primordial non-Gaussianity, especially making use of the multi-tracer
      technique \cite{2014MNRAS.442.2511F}. Our aim here is to add on this result by
      considering the different cross-correlations with intensity mapping, as well as the
      potential of these cross correlations to detect relativistic effects.
      
      Our models for the redshift distribution and bias functions for the different
      radio sources considered here follow the method discussed in Appendix B of
      \cite{2015arXiv150507596A}, based on the measurements of the luminosity functions
      by \cite{2001ApJ...554..803Y,2003ApJ...598..886U,2001MNRAS.322..536W}, used in
      \cite{2008MNRAS.388.1335W} to produce semi-empirical simulations for radio surveys.
      We will consider here four different galaxy populations: normal star-forming
      galaxies, starbursts, radio-quiet AGN and radio-loud AGN of the FR-I type. Owing
      to their very small number density, we omitted FRII galaxies from this analysis.
      Note that, as described in \cite{2014MNRAS.442.2511F}, separating the combined
      sample of radio galaxies into these four different types could be a challenging
      task, especially given the enormous size of the sample that the SKA will be able
      to collect. Furthermore, although it should be possible to distinguish radio-loud
      AGN from other sources on a morphological basis, differentiating radio-quiet quasars
      would probably require X-ray observations with a different experiment. Further
      separating normal star-forming galaxies from starbursts would require optical and
      redshift information, which would be difficult to obtain for the whole redshift
      range. For this reason we produce our forecasts for two cases: a pessimistic 
      scenario with no source differentiation (i.e. one single tracer) and an optimistic
      one with two disjoint samples, given by radio-loud AGN on the one hand and a
      combination of star-forming galaxies, starbursts and radio-quiet quasars on the
      other. We will call these two samples ``sample 2'' and ``sample 1'' respectively.
      
      As for intensity mapping, we consider here an SKA1-MID configuration, which should be
      able to detect radio sources to $z\sim5$ over three quarters of the sky by
      integrating their continuum flux in the band 350-1050 MHz with an rms noise of
      $S_{\rm rms}=1\uJy$, and we will assume a $5\sigma$ detection threshold (i.e.
      $S_{\rm cut}=5\uJy$). The redshift distribution of the two populations considered
      here as well as the combined sample for this flux cut can be seen in Figure
      \ref{fig:nz_continuum}, and the corresponding clustering, magnification and evolution
      biases are shown in Figure \ref{fig:bias_functions}, together with those for HI
      intensity mapping.
      
      The noise power spectrum for continuum surveys is simply given by Poisson shot-noise
      $N_\ell^{\rm cont}=\bar{N}_\Omega^{-1}$, where $\bar{N}_\Omega$ is the angular number
      density of sources (per steradian) in the survey. We must note that the intensity
      mapping noise is not completely uncorrelated with that of continuum sources. At low
      redshifts most of the HI is located inside galaxies, and so many of the galaxies
      contributing to the HI line emission will also be detected in the continuum survey.
      However, the contribution of this Poisson term to the total noise power spectrum is
      very small compared to the instrumental noise levels \cite{2015ApJ...803...21B},
      and therefore we neglect these correlations.
  
  \subsection{Forecasts}\label{ssec:radio_forecasts}
    \begin{figure}
      \begin{center}
        \includegraphics[width=0.49\textwidth]{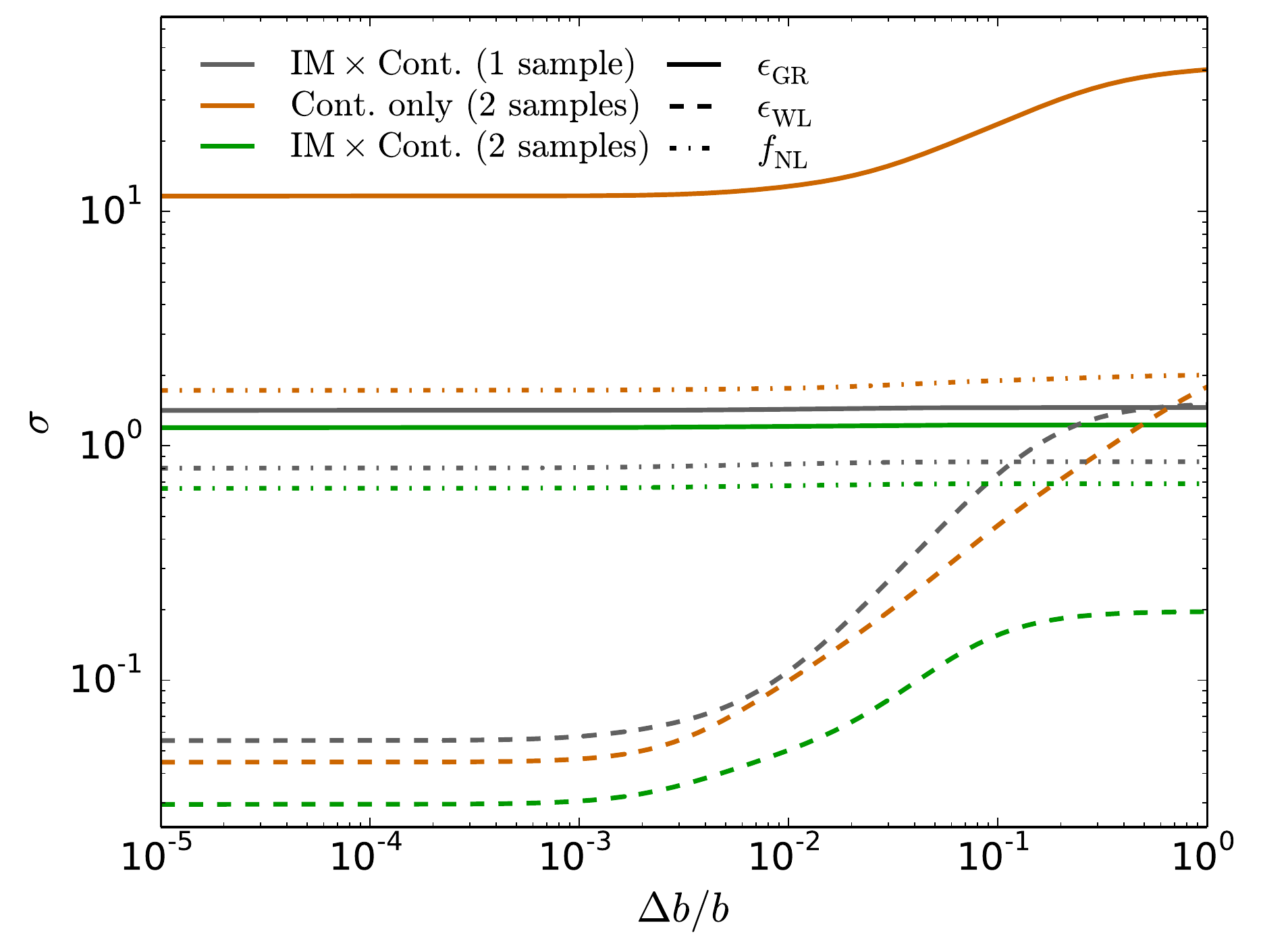}
        \caption{Dependence of the constraints on $\fnl$, $\egr$ and
	         $\ewl$ on a constant Gaussian prior on the clustering bias
	         $b(z)$. The constraints on $\ewl$ exhibit a much stronger
	         dependence on $\Delta b$ due to the fact that lensing magnification
	         can only be constrained through the continuum survey.}
       \label{fig:sigma_bz_im_cont_05}
      \end{center}
    \end{figure}
    \begin{figure}
      \begin{center}
        \includegraphics[width=0.49\textwidth]{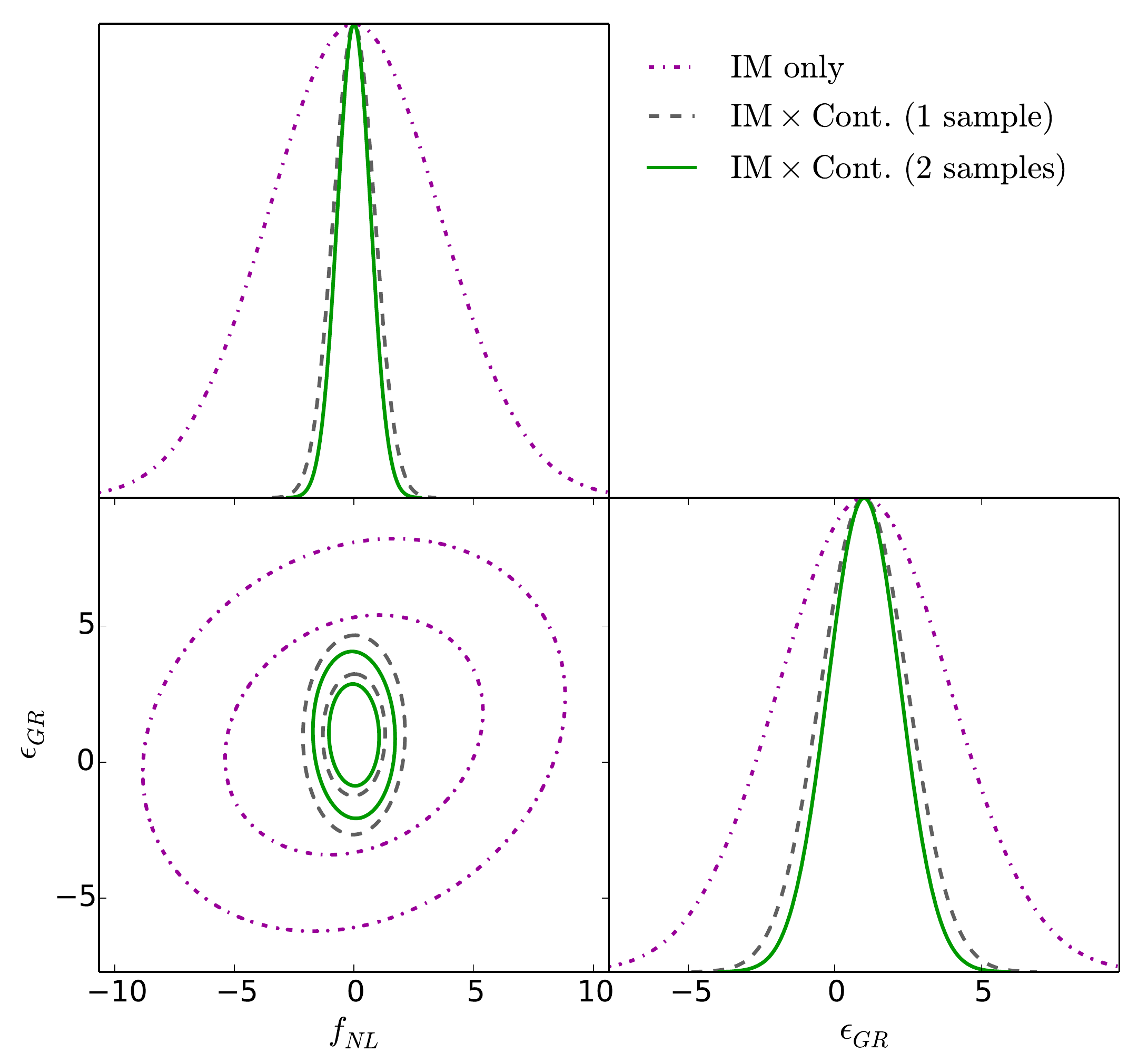}
        \caption{Joint constraints for $\fnl$ and $\egr$ for the best three combinations
                 of radio experiments: intensity mapping-only (purple), intensity mapping
                 combined with a single continuum sample (grey) and intensity mapping
                 combined with the two different continuum samples (green). Although the
                 GR effects are not detectable, the improvement in the uncertainty on
                 $\fnl$ due to the addition of continuum data is remarkable.}
       \label{fig:ellipses_im_cont_05}
      \end{center}
    \end{figure}
    We can expect the advantages of combining intensity mapping and continuum observations
    in a multi-tracer sense to be threefold: first, the large clustering bias of
    e.g. AGN combined with the very low $b$ of neutral hydrogen should significantly boost
    the amplitude of the $\fnl$ signal. Secondly, the absence of perturbations on transverse
    scales for intensity mapping should help to separate the GR effects and lensing
    magnification contributions in the cross-correlation. Finally, the main drawback
    of continuum surveys is the absolute lack of radial information, and therefore, given
    the good redshift resolution of intensity mapping, combining both datasets should 
    significantly improve any constraints.
    
    We have computed the forecasted errors on $\fnl$, $\egr$ and $\ewl$ for four
    different tracer combinations:
    \begin{itemize}
     \item Case A: using only intensity mapping (IM only). These results were already
           presented in \cite{2015arXiv150507596A,2013PhRvL.111q1302C}.
     \item Case B: combining intensity mapping and the joint continuum sample
           (IM$\times$Cont. (1 sample)).
     \item Case C: combining intensity mapping and the two separated radio populations
           (IM$\times$Cont. (2 samples)).
     \item Case D: combining the two radio populations (Cont. only
           (2 samples)).
    \end{itemize}
    In all cases we assumed complete overlap between the intensity mapping and continuum
    surveys (i.e. $f_{\rm sky}=0.75$).
    
    As in the case of LSST, we first look at the feasibility of detecting GR corrections by computing
    the uncertainty on $\egr$ in the most optimistic case, with all parameters, except for
    $\fnl$, fixed:
    \begin{align}\nonumber
      \sigma_{\rm A}(\egr)=2.75,\hspace{12pt}
      \sigma_{\rm B}(\egr)=1.40,\\\nonumber
      \sigma_{\rm C}(\egr)=1.19,\hspace{12pt}
      \sigma_{\rm D}(\egr)=10.4.
    \end{align}
    Thus, even in the best-case scenario it would be impossible to observe the effect of
    the relativistic corrections above $1\sigma$. Following the argument laid out in
    section \ref{ssec:multi_why} it is easy to understand why. First of all, the
    signal-to-noise level for both types of experiment is significantly lower than for
    photometric surveys. This is due to the smaller number density of radio sources above
    $5\uJy$ as well as the the high noise level expected for intensity mapping. Secondly,
    unlike in the case of red and blue galaxies, the magnification and evolution biases
    of the different tracers are not that different from one another.
    
    As was reported in \cite{2015arXiv150507596A}, the lack of radial information for
    continuum surveys makes the constraints on our large-scale observables very sensitive
    to the priors assumed on any nuisance or cosmological parameter. We can expect that the
    cross-correlation with intensity mapping should help in breaking some of these degeneracies,
    and therefore we have studied them in detail. We obtain results that are mostly
    equivalent to those found for LSST: the constraints on $\egr$ and $\fnl$ are insensitive
    to the prior on the clustering bias, but show a clear dependence on $\Delta\fevo$,
    although in this case they can be optimized with a looser prior of $\Delta\fevo\sim1$.
    The constraints on the amplitude of the lensing magnification term $\ewl$, on the
    other hand, are significantly more sensitive to the priors on $b(z)$ and $s(z)$ than
    in the case of LSST. The main reason for this is that we can only constrain $\ewl$
    through the continuum data, since the lensing term cancels exactly to first order for
    intensity mapping, and thus the lensing amplitude is more degenerate with other
    parameters. We show the dependence of $\ewl$ on $\Delta b$ in Fig.
    \ref{fig:sigma_bz_im_cont_05}.

    As we did for LSST, we define an optimistic and pessimistic set of priors, given
    by $\Delta s=\Delta\fevo=0.1$ and $\Delta s=\Delta\fevo=1$ respectively. In both
    cases we will assume that the clustering bias can be constrained to a 10\% precision
    ($\Delta b/b=0.1$) based on smaller-scale data \cite{2014MNRAS.440.2322L}.
    
    Our final constraints on $\fnl$ and $\egr$ are shown jointly in Figure
    \ref{fig:ellipses_im_cont_05} for optimistic priors and listed in Table
    \ref{tbl:constraints_summary}. Although the GR corrections remain unobservable for
    radio experiments, the improvement of the constraints on $\fnl$ due to combining
    intensity mapping and continuum data is remarkable.

\section{Synergies}\label{sec:im_vs_photo}
  \begin{figure}
    \begin{center}
      \includegraphics[width=0.49\textwidth]{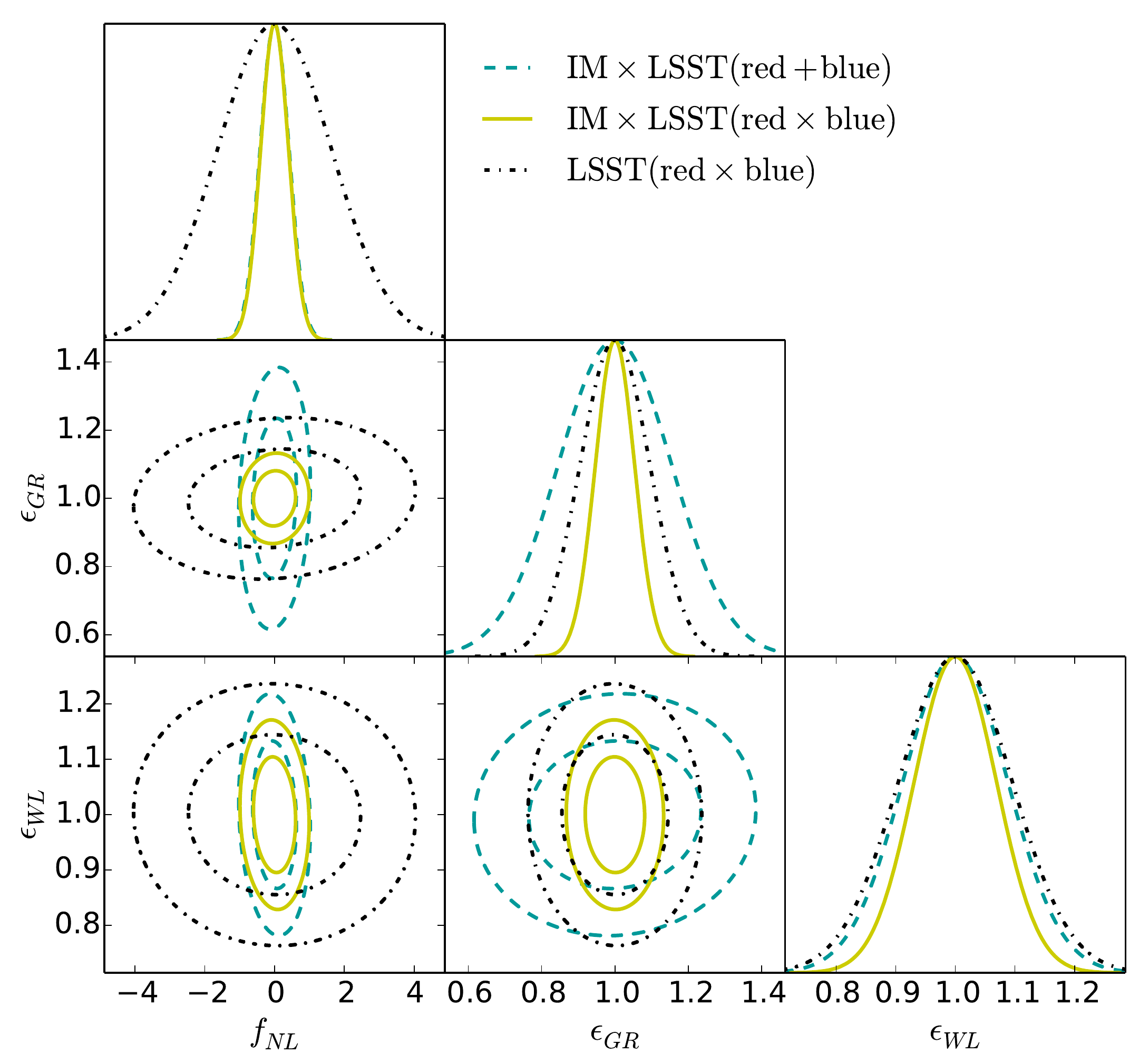}
      \caption{Joint constraints on $\fnl$, $\egr$ and $\ewl$ for the combination
               of both red and blue galaxies in LSST (black ellipses), for a
               combination of SKA1-MID intensity mapping with the full LSST
               sample (cyan) and for intensity mapping combined with both red
               and blue galaxies for (yellow).}
     \label{fig:ellipses_ska_x_lsst}
    \end{center}
  \end{figure}
  In the last two sections we have described the potential of two key next-generation
  experiments, SKA-MID and the LSST, to constrain cosmological observables on ultra-large
  scales, considering the potential multi-tracer analyses that could be performed using
  the capabilities of each experiment independently. We would now like to explore the
  possibility of combining the tracers probed by both surveys jointly, and the potential
  improvement that this synergy would represent over the results presented above.

  Both experiments will be located at similar latitudes and will therefore cover almost
  equivalent regions of the sky. There should also be a significant redshift overlap
  between the LSST photometric survey and the SKA-MID HI intensity mapping experiment.
  The number of potential benefits of cross-correlating the datasets from both experiments
  is therefore immense, e.g.: the ability of LSST to provide approximate redshifts
  for many of the sources in the SKA continuum survey, the ability of the SKA to provide
  more accurate redshift information through its HI survey to improve the LSST photo-$z$
  calibration, or the invaluable cross-checks that will be performed to mitigate the 
  effects of survey-specific systematic uncertainties (see \cite{2015aska.confE.145B}
  for a recent review of potential synergies). Since such potential benefits can not be
  ignored, this synergy will surely take place, and therefore it makes sense to forecast
  for the potential impact on ultra large-scale cosmology.

  We will present our forecasts in two cases:
  \begin{itemize}
    \item Combining SKA-IM and the LSST full sample, without differentiating between red and
      blue galaxies.
    \item Combining SKA-IM and the two red and blue LSST samples (i.e. three tracers in total).
  \end{itemize}
  We do not consider cross-correlations between LSST and the SKA continuum survey
  for two reasons. Firstly, a large fraction of the radio galaxies detected by the
  SKA should be present in the LSST dataset too. Therefore these are not completely
  disjoint tracers, and evaluating their noise cross-power spectrum, crucial for the
  multi-tracer method, would be very uncertain. Secondly, the benefits of combining
  IM with a photometric survey are clearer: an improved coverage of the small 
  radial scales due to the better spectral resolution of intensity mapping and an
  improved coverage of small transverse scales due to the better angular resolution
  of optical surveys. Finally, as we discussed in section \ref{sssec:im}, the
  low clustering bias, mild evolution bias and the absence of linear-order perturbations
  in the transverse distance hamper the ability of intensity mapping to constrain
  $\fnl$ and $\egr$ on its own. However, those same characteristics make it a very
  useful probe to combine with other tracers in the multi-tracer scheme.

  Although both experiments should cover almost the same area of the sky, we can
  expects a certain loss of overlap caused by regions of the sky dominated by different
  systematics (e.g. dust extinction for LSST and galactic synchrotron for SKA-IM). The
  forecasts presented here conservatively assume that both experiments will yield
  useable observations in a patch of area $f_{\rm sky}=0.4$. Note that we treated this
  value as the sky fraction in which both experiments can be cross-correlated, however
  we still use the full SKA area ($f_{\rm sky}=0.75$) when computing the noise power
  spectrum for intensity mapping.
    
  The full LSST sample was modelled by combining the red and blue luminosity functions,
  and conservatively assuming a photo-$z$ uncertainty $\sigma_0=0.05$. Following the
  prescription outlined in Section \ref{ssec:photo_observables}, this entails dividing
  the sample into the same 9 redshift bins used for the blue population.
 
  The joint constraints on $\fnl$, $\ewl$ and $\egr$ for the two cases above, together
  with the constraints found for LSST alone are shown in Figure
  \ref{fig:ellipses_ska_x_lsst}, where an optimistic prior $\Delta s=\Delta\fevo=0.1$
  was used. The quantitative results are listed in Table \ref{tbl:constraints_summary}
  for both pessimistic and optimistic priors. Several conclusions can be drawn from this
  exercise: 
  \begin{itemize}
    \item In spite of the low clustering bias of neutral hydrogen, the errors on
          $\fnl$ improve immensely by combining LSST and intensity mapping
          (by a factor of $\sim3-4$ compared to the LSST-only case).
    \item Although we observed that the large evolution bias of red galaxies was crucial
          in boosting the detection level of GR corrections when jointly analyzing
          the red and blue samples for LSST, we now see that the cross-correlation of
          HI intensity mapping and the full LSST sample, which does not display such
          a large evolution bias, could still yield a measurement of $\egr$ above
          the $5\sigma$ level. In the most optimistic case (optimistic priors and
          a joint analysis of the three samples) this increases to $\sim20\sigma$.
    \item Due to the cancellation of the lensing term for intensity mapping the
          improvement in the detection of the large-scale lensing magnification is
          only mild.
  \end{itemize}
  \begin{table*}[t]
    \centering{
      {\renewcommand{\arraystretch}{1.3}
      \begin{tabular}{|l|l|c|c|c|}
        \hline
        {\sf Experiment type} & {\sf Tracers} 
        & $\sigma(\fnl)$ & $\sigma(\egr)$ & $\sigma(\ewl)$ \\
        \hline
        Photometric survey & LSST, red-only                     
        & ~~4.53 (4.54)~~ & ~~1.65 (1.70)~~ & ~~0.18 (0.67)~~ \\
        (LSST)             & LSST, blue-only                    
        & ~~1.71 (1.72)~~ & ~~3.45 (3.48)~~ & ~~0.12 (0.61)~~ \\
                           & LSST, red $\times$ blue            
        & ~~1.62 (1.63)~~ & ~~0.10 (0.17)~~ & ~~0.10 (0.36)~~ \\
                           & DES, red $\times$ blue             
        & ~~7.18 (7.20)~~ & ~~0.29 (0.32)~~ & ~~0.06 (0.06) $(^*)$~~ \\
        \hline
        Radio      & IM-only                                      
        & ~~3.00 (3.01)~~ & ~~2.71 (2.75)~~ & ---        \\
        (SKA1-MID) & IM$\times$Cont., 1 sample  
        & ~~0.86 (0.89)~~ & ~~1.47 (1.95)~~ & ~~1.16 (3.80)~~ \\
                   & IM$\times$Cont., 2 samples 
        & ~~0.69 (0.71)~~ & ~~1.23 (1.43)~~ & ~~0.33 (2.21)~~ \\
                   & Continuum-only, 2 samples
        & ~~1.91 (1.97)~~ & ~~23.9 (31.1)~~ & ~~0.56 (2.85)~~ \\
        \hline
        Synergy                & IM$\times$all            
        & ~~0.41 (0.41)~~ & ~~0.15 (0.39)~~ & ~~0.09 (0.32)~~ \\
        (SKA1-MID$\times$LSST) & IM$\times$red$\times$blue      
        & ~~0.40 (0.40)~~ & ~~0.05 (0.12)~~ & ~~0.07 (0.19)~~ \\
        \hline
      \end{tabular}
      }
    }
    \caption{Forecasted constraints on $\fnl$, $\egr$, and $\ewl$ for the
      different experiments explored in this work. In each case, the constraints
      assuming optimistic and pessimistic priors on the magnification and evolution
      biases are shown with and without parentheses respectively. $(^*)$Note that the
      uncertainty on $\ewl$ is shown to be smaller for DES than LSST. This
      does not mean that DES would be better at measuring lensing, only that
      the value of the magnification bias at the magnitude limit of DES
      ($\sim24$) enhances the lensing magnification effect. LSST would
      be able to greatly improve on this result by adopting a different
      (lower) flux cut. This was already noted in \cite{2015arXiv150601369M}.}
    \label{tbl:constraints_summary}
  \end{table*}

\section{Discussion}\label{sec:discussion}
  In this paper we have shown how more than one tracer of the cosmic density
  field can be used to extract precise measurements of the ultra large-scale
  features of the universe. We have shown that a judicious combination of
  photometric redshift surveys with cosmological radio surveys will allow us
  to obtain precise constraints of primordial non-gaussianity, of weak
  lensing on large scales and most significantly, of general-relativistic
  effects which depend on various combinations of the gravitational potential.
  Our final results are summarized in Table \ref{tbl:constraints_summary}
  in terms of the marginalized uncertainties on $\fnl$ and  the amplitudes
  of the GR corrections ($\egr$) and lensing magnification ($\ewl$).

  More specifically, we have shown that the best way to optimize the detection
  of GR effects is to use a combination of low-noise tracers (i.e. high number
  density) with very different evolution and magnification biases. We find that
  ongoing and future photometric redshift surveys will be able to satisfy both
  conditions, given the large expected number density of observed sources and
  the possibility of separating the galaxy sample into two populations, blue and
  red galaxies, with different evolution of their luminosity functions. Indeed
  we have seen how the rapid decay in the number density of red galaxies above
  redshift $\sim1$ translates itself into a large $\fevo$ that boosts the signal
  of the GR corrections and could enable LSST to detect them with a significance
  of $\sim5-10\,\sigma$. Using the same method we have also shown that DES might
  be able to make a lower significance detection ($\sim3\,\sigma$) in a shorter
  term.

  We have also explored the possibility of combining two of the main cosmological
  radio probes, HI intensity mapping and continuum surveys, planned for the
  Square Kilometre Array. We have seen that in this case, the higher noise levels
  make a detection of the GR corrections unviable, in spite of the larger volume
  covered in comparison with LSST. However, the contrast between the low clustering
  bias of neutral hydrogen and its large value for certain radio sources makes the
  combined analysis of both probes very beneficial to reduce the uncertainties
  on the level of primordial non-Gaussianity. We have seen that, in the worst-case
  scenario (1 single continuum sample and pessimistic priors), the error on $\fnl$
  could improve by a factor of 3 with respect to the constraints from intensity
  mapping alone.
  
  Finally, since LSST and SKA will carry out their observations over a very similar
  region of the sky, the benefits of cross-correlating both experiments will be immense
  for many science cases, and here we have studied the potential impact of this
  synergy in the study of cosmology on ultra-large scales. We have shown that 
  it would be possible to improve the constraints on $\fnl$ by a
  factor of at least $2-3$ with respect to the individual constraints from each
  experiment. The detection level of the GR effects would also improve significantly
  and, more interestingly, the combination of the full LSST sample and SKA1 intensity
  mapping should also be able to detect these effects above $5\sigma$ regardless of
  the large evolution bias of red galaxies.

  We have shown that the detectability of these new effects is an exciting prospect, yet many
  important challenges must be overcome before such measurements can be carried
  out. Firstly, the estimates of clustering statistics on ultra-large scales in
  most experiments are riddled with systematic uncertainties. Intensity mapping
  is a perfect example of this: galactic and extragalactic foregrounds
  have radio intensities that are several orders of magnitude larger than the cosmological
  HI signal. Although both components can in principle be separated based on their
  very different spectral behaviour, any foreground residual will necessarily contaminate
  the signal on large scales, both angular (due to the large-scale features of galactic
  foregrounds) and radial (due to their smooth frequency dependence). As shown in
  \cite{2015MNRAS.447..400A}, imperfect foreground cleaning can lead to the cross-correlations
  between well separated frequency bins being unreliable, and, as was reported in
  \cite{2015arXiv150507596A}, excluding these cross-correlations from the analysis
  can have a very detrimental effect on the uncertainty on large-scale parameters such
  as $\fnl$. Other instrumental issues, such as the effect of correlated $1/f$ noise
  in the data, a bad characterization of the instrument's beam or the contamination
  from polarized foregrounds due to polarization leakage can introduce even more troublesome
  systematics in the data that must be correctly minimized in order to make these measurements
  possible.
  \begin{figure}
    \begin{center}
      \includegraphics[width=0.49\textwidth]{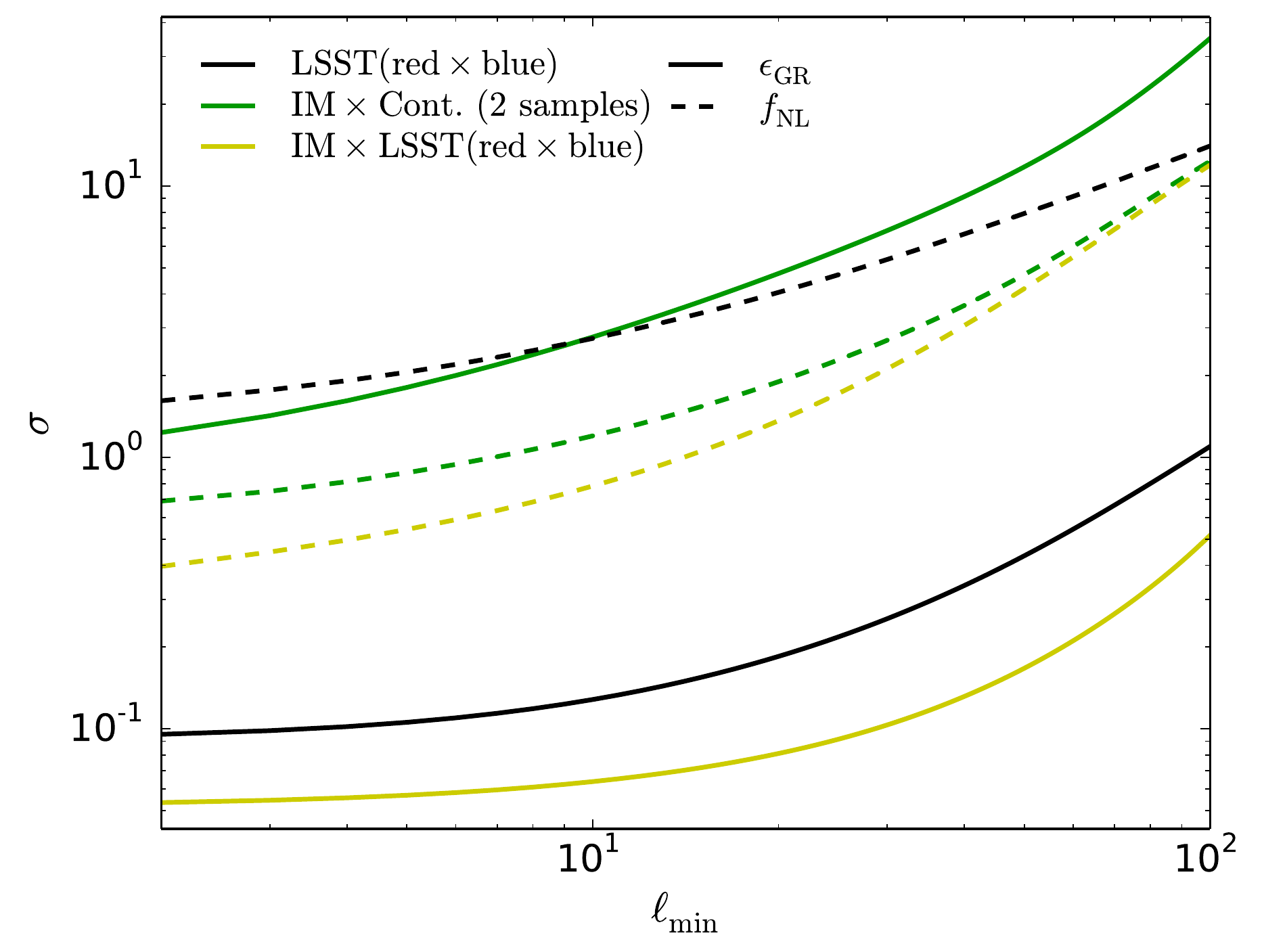}
      \caption{Dependence of the constraints on $\egr$ (solid) and $\fnl$ (dashed) on the
               largest angular scale included in the analysis (given by the minimum multipole
               $\ell_{\rm min}$), for three different tracer combinations: LSST red and blue
               galaxies (black), intensity mapping and our 2 continuum samples (green) and IM
               combined with both LSST samples (yellow). The constraints degrade significantly
               after $\ell_{\rm min}\sim10$.}
     \label{fig:sigma_lmin_best}
    \end{center}
  \end{figure}
  
  Optical surveys are also subject to a sizeable collection of systematics. Effects such as
  galactic dust extinction and airmass can modify the observed magnitude and generate a
  non-uniform magnitude limited sample which can modify the observed clustering statistics
  on large scales. Star contamination, i.e. the accidental inclusion of stars in the galaxy
  sample, and star obscuration, as reported in \cite{Ross:2011cz}, can also have a
  significant impact on the clustering observed on large angular distances, since both
  effects grow towards the galactic plane. Another specific source of systematic uncertainty
  for photometric surveys is the incorrect characterization of the window functions
  corresponding to each redshift bin. This can be caused by uncertainties in the galaxy
  redshift distribution, due to the absence of precise redshifts, and to inaccuracies in
  the characterization of the photo-$z$ distribution as a function of redshift.
  
  Most of these systematic effects, unless corrected for, will limit our ability to use the
  observations on the largest angular scales. We have studied the dependence of our forecasts
  on the largest scale included in the analysis (i.e. the minimum angular multipole
  $\ell_{\rm min}$) in Figure \ref{fig:sigma_lmin_best} for our best three tracer
  combinations (LSST red and blue, IM combined with the two continuum samples and IM
  combined with the two LSST samples). The constraints on $\fnl$ and $\egr$ degrade
  significantly above $\ell_{\rm min}\sim10$.
  
  In addition to these systematic effects, we have also identified other challenges that
  must be tackled before this detection can be achieved. We have found that having at least
  two tracers with very different magnification and/or evolution biases is key to boosting
  the signal of GR effects, and we have illustrated this point with the case of red and
  blue galaxies in photometric surveys. However, we have also seen that, even if such tracers
  can be found, a sufficiently precise knowledge of their joint magnitude-redshift distribution
  is necessary to statistically detect the GR effects, which would otherwise be swamped by the
  uncertainties on $s(z)$ and $\fevo$. In the case of LSST we have shown that priors
  of the order of $\Delta s\simeq\Delta\fevo\simeq0.1$ would be needed to optimize this
  measurement. To put these uncertainties into perspective, we should note that the
  uncertainty in the value of $\fevo$ propagated from the errors in the
  estimate of the luminosity function for red galaxies made by \cite{2007ApJ...665..265F},
  based on data from the DEEP2 and COMBO-17 surveys, is of the order of $\Delta\fevo\sim2$
  at high redshifts ($z\gtrsim0.8$), where the GR effects are maximally boosted.
  Assuming that this uncertainty decreases with the square-root of the observed number
  of galaxies, spectroscopic observations of a representative sample of $\sim1.5$ million
  galaxies with redshifts above $\sim0.8$ would be needed to achieve the optimal
  sensitivity. This would correspond to a fraction of $\sim2\%$ of the total red sample
  expected for LSST in the same redshift range. The Dark Energy Spectroscopic Instrument
  (DESI) \citep{2013arXiv1308.0847L}, for instance, should be able to measure spectra for
  $\sim4$ million luminous red galaxies in the redshift range $0.4\lesssim z\lesssim1.0$,
  although these observations would be carried out in the northern hemisphere. 
  
  Finally, it is worth noting that the detection levels and uncertainties reported here
  depend critically on the fiducial models used for the bias functions of the tracers
  under study. Although we have tried to use reliable and conservative assumptions for
  these models, these quantities are still uncertain, and therefore the quantitative
  results could change as those uncertainties are reduced. On the other hand, this
  also implies that these results could actually be improved if one could find more
  optimal tracer classifications (e.g. beyond the traditional red-blue split) enhancing
  the amplitude of the relativistic corrections.
  
  If we can tackle all these challenges, then, by measuring ultra large-scale modes in the
  density field, we will be able to break out of what has been the conventional arena of
  large scale structure studies - the quasi-static regime - on substantially sub-horizon
  scales. Until now, most surveys of large-scale structure (such as redshift or
  weak lensing surveys) have probed what can be thought of as the Newtonian
  regime of cosmological perturbation theory, where the time dependence of the
  gravitational potentials can be ignored \cite{Noller:2013wca}. Now, by probing scales
  which span the cosmological horizon, this time dependence comes into play, giving us a
  completely new window on the evolution of cosmic perturbations.

  Accessing ultra-large scales can be of particular significance in constraining
  modifications to General Relativity \cite{2013PhRvD..87f4026H}. It has been shown
  that in the quasi-static regime, cosmological data will only constrain a limited subspace
  of possible modifications to gravity \cite{Baker:2013hia,Baker:2014zva,Leonard:2015hha}.
  To extend the scope of the constraints, it is essential to break any degeneracies
  that arise on small scales by using the time variation of the gravitational
  potentials on large scales. As shown by \cite{BakerBull,Lombriser:2015cla} one
  will expect specific signatures (of order of a few percent) which, with the
  methods proposed in this paper, might be observable with future surveys.

  Given that we are targeting ultra-large scales, one should bear in mind the
  mild evidence for deviations from the standard, $\Lambda$CDM, concordance model
  on large scales. The recent Planck measurements of the the CMB on large scales has
  found a number of anomalies - suppression of power on very large angles, correlations
  between the large angle multipoles and an anomalously  cold spot - which have yet to
  be understood \cite{Ade:2015sjc}. Cosmic variance severely restricts any attempts at better
  characterizing these anomalies with the CMB alone. One would hope, however, that with the
  possibility of better constraining large scale modes with multiple tracers, one could
  substantially improve our characterisation of these large scale anomalies with complementary
  constraints.

  We have focused on extracting information at the level of auto and cross-correlations
  of the various tracers. An interesting, uncharted, avenue would be to attempt to use
  these techniques to extract {\it maps} of each of the effects that come into play. In
  other words, and much like what is done in the case of the CMB \cite{Crittenden:1995ak},
  one could imagine constructing linear combinations of the various observables (which are
  three dimensional maps of a particular tracer) to extract, for example, maps of the
  gravitational potential, their derivatives, large scale flows, etc. This is a somewhat
  ambitious goal but if achievable would open up a completely new window into the large
  scale properties of space-time and matter; a direct comparison between the different
  maps, and a comparison between these maps and other cosmic observables might be used
  to uncover different cosmological properties which are usually buried in the standard
  correlation function (or power spectrum) analysis.

  Finally, we have focused on a particular choice of current and future surveys -
  photometric redshift surveys and cosmological radio surveys - but this does not cover
  all possible observational approaches. Spectroscopic galaxy redshift surveys, such as
  Euclid \cite{Laureijs:2011gra} and WFIRST \cite{Green:2012mj}, will also attempt to
  constrain the effects we discuss in this paper, albeit on smaller scales and with a
  lower source density. A more speculative, ambitious and, in our mind, exciting approach
  using an all sky, wide band survey, SPHEREX, would have all the right characteristics
  for the methods proposed here. As argued in \cite{Dore:2014cca}, the number density and
  bias of these sources should be sufficiently large to aid, and complement, the
  constraints that will be found with the combination of probes found here. In addition to
  these probes of galaxy clustering, it would also be worth exploring the potential benefits
  of combining them with other cosmological observables, such as weak lensing, CMB lensing
  or even the CMB itself. These combinations could benefit from the particular properties of
  these tracers (e.g. zero bias for weak lensing and minimal clustering variance for the CMB),
  significantly improving the constraints reported here. We are clearly entering an era in
  which precision measurement of ultra large-scale modes can become a core science goal of
  observational cosmology. 

\section*{Acknowledgments}
  We thank Phil Bull, Elisa Chisari, Jo Dunkley, Ruth Durrer, Roy Maartens, Francesco
  Montanari and Mario Santos for useful comments and discussions. DA is supported by ERC
  grant 259505. PGF acknowledges support from STFC, BIPAC and the Oxford Martin School.
 
\bibliography{paper}

\appendix
\section{The linear order components of the fluctuation in source number counts}
  \label{app:full_expressions}
  The complete expressions, to linear order, for the transfer function of the perturbation
  in the source number counts used in Equation \ref{eq:th_cl} is given by
  \cite{2013JCAP...11..044D}:
  \begin{widetext}
    \begin{align}\label{eq:terms0}
      &\Delta^{{\rm D},i}_\ell(k)\equiv\int d\eta \,b\,\tilde{W}_i\,
      \delta_{M,{\rm syn}}(k,\eta)\,j_\ell(k\chi(\eta)),\hspace{12pt}
      ~~~~~~~~\Delta^{{\rm RSD},i}_\ell(k)\equiv\int d\eta\,(aH)^{-1}
      \tilde{W}_i(\eta)\,\theta(k,\eta)\,
      j_\ell''(k\chi(\eta)),\\
      &\Delta^{{\rm L},i}_\ell(k)\equiv \ell(\ell+1) \int d\eta\,
      \tilde{W}^{\rm L}_i(\eta)\,
      (\phi+\psi)(k,\eta)\,j_\ell(k\chi(\eta)),\hspace{12pt}
      \Delta^{{\rm V1},i}_\ell(k)\equiv \int d\eta\,(\fevo-3)\,aH\,\tilde{W}_i(\eta)\,
      \frac{\theta(k,\eta)}{k^2}\,j_\ell(k\chi(\eta)),\\
      &\Delta^{{\rm V2},i}_\ell(k)\equiv \int d\eta\,
      \left(1+\frac{H'}{aH^2}+\frac{2-5s}{\chi\,aH}+5s-\fevo\right)
      \tilde{W}_i(\eta)\,\frac{\theta(k,\eta)}{k}\,j_\ell'(k\chi(\eta)),\\
      &\Delta^{{\rm P1},i}_\ell(k)\equiv \int d\eta\,
      \left(2+\frac{H'}{aH^2}+\frac{2-5s}{\chi\,aH}+5s-\fevo\right)
      \tilde{W}_i(\eta)\,\psi(k,\eta)\,j_\ell(k\chi(\eta)),\\
      &\Delta^{{\rm P2},i}_\ell(k)\equiv \int d\eta\,
      (5s-2)\tilde{W}_i(\eta)\,\phi(k,\eta)\,j_\ell(k\chi(\eta)),\hspace{12pt}
      \Delta^{{\rm P3},i}_\ell(k)\equiv \int d\eta\,
      (aH)^{-1}\tilde{W}_i(\eta)\,\phi'(k,\eta)\,j_\ell(k\chi(\eta)),\\\label{eq:isw}
      &\Delta^{{\rm P4},i}_\ell(k)\equiv \int d\eta\,
      \tilde{W}^{\rm P4}_i(\eta)\,(\phi+\psi)(k,\eta)\,j_\ell(k\chi(\eta)),\hspace{12pt}
      \Delta^{{\rm ISW},i}_\ell(k)\equiv \int d\eta\,
      \tilde{W}^{\rm ISW}_i(\eta)\,(\phi+\psi)'(k,\eta)\,j_\ell(k\chi(\eta)),
    \end{align}
    where $j_\ell(x)$ is the spherical Bessel function of order $\ell$ and where we have
    defined the window functions
    \begin{align*}\label{eq:windows}
      &\tilde{W}_i(\eta(z))\equiv W_i(z)\left(\frac{d\eta}{dz}\right)^{-1},\hspace{12pt}
      \tilde{W}^{\rm L}_i(\eta)\equiv\int_0^\eta d\eta'\tilde{W}_i(\eta')
      \frac{2-5s(\eta')}{2}\frac{\chi(\eta)-\chi(\eta')}{\chi(\eta)\chi(\eta')},\\
      &\tilde{W}^{\rm P4}_i(\eta)\equiv\int_0^\eta d\eta'\tilde{W}_i(\eta')
      \frac{2-5s}{\chi},\hspace{12pt}
      \tilde{W}^{\rm ISW}_i(\eta)\equiv\int_0^\eta d\eta'\tilde{W}_i(\eta')
      \left(1+\frac{H'}{aH^2}+\frac{2-5s}{\chi\,aH}+5s-\fevo\right)_{\eta'}.
    \end{align*}
  \end{widetext}
  The quantities $\delta_{M,{\rm syn}}$, $\theta$, $\phi$ and $\psi$ in the equations above
  are respectively: the transfer function for the matter density perturbations in the
  comoving synchronous gauge, the transfer function for the peculiar velocity divergence
  in the conformal Newtonian gauge and transfer functions for the two metric potentials
  in the same gauge, defined by the line element:
  \begin{equation}
   ds^2=-a^2(\eta)\,\left[(1+2\psi)d\eta^2-(1-2\phi)\delta_{ij}dx^idx^j\right].
  \end{equation}

\end{document}